%% file: main.tex
\documentclass[journal]{IEEEtran}
\usepackage{amsmath,amsfonts}
\usepackage{bm}
\usepackage{algorithmic}
\usepackage{algorithm}
\usepackage{array}
\usepackage[caption=false,font=normalsize,labelfont=sf,textfont=sf]{subfig}
\usepackage{textcomp}
\usepackage{stfloats}
\usepackage{url}
\usepackage{verbatim}
\usepackage{graphicx}
\usepackage{cite}
\usepackage{xcolor}
\usepackage{colortbl}
\usepackage{pifont}
\usepackage{capt-of}
\usepackage{hyperref}
\usepackage{booktabs}
\usepackage{makecell}
\usepackage{multirow}

\begin{document}

\title{Benchmarking Neural Speech Compression from a Rate-Distortion Perspective}

\author{Jun Xu\IEEEauthorrefmark{1}, Zhengxue Cheng\IEEEauthorrefmark{1},~\IEEEmembership{Member,~IEEE,} Fengxi Zhang, Yuhan Liu, \\ Li Song$^{\dagger}$,~\IEEEmembership{Senior Member,~IEEE,} 
and Wenjun Zhang,~\IEEEmembership{Fellow,~IEEE}
\thanks{\IEEEauthorrefmark{1}Jun Xu and Zhengxue Cheng contributed equally to this work.}
\thanks{$^{\dagger}$Corresponding author: Li Song.}
\thanks{Jun Xu, Zhengxue Cheng, Fengxi Zhang, Yuhan Liu, Li Song, and Wenjun Zhang are with the School of Information Science and Electronic Engineering, Shanghai Jiao
Tong University, Shanghai 200240, China (e-mail: \{xujunzz, zxcheng, zhangfengxi, liu1025221459, song\_li,
zhangwenjun\}@sjtu.edu.cn)
}
}

\markboth{Journal of \LaTeX\ Class Files,~Vol.~14, No.~8, August~2021}%
{Shell \MakeLowercase{\textit{et al.}}: A Sample Article Using IEEEtran.cls for IEEE Journals}


\maketitle

\begin{abstract}
Learning-based speech compression has achieved promising low-bitrate performance, but many neural speech codecs still describe quantized latents with preset-rate discrete symbols or apply entropy coding only after symbol generation.
Such designs decouple representation learning from probability modeling, limiting their ability to exploit the non-uniform usage and temporal dependencies of learned speech latents.
In this paper, we benchmark neural speech compression from a rate--distortion perspective and further investigate entropy-constrained coding for low-bitrate speech compression.
We first formulate a unified learning-based speech coding pipeline and provide a benchmark-style analysis of recent neural speech codecs, showing that explicit probability modeling remains underexplored in learned speech compression.
We then propose ECC, an Entropy-Constrained Codec that combines scalar quantization with a learned entropy model. ECC integrates hyperprior-based side information, channel-wise context modeling, latent residual prediction, and lightweight temporal modeling to estimate latent likelihoods for rate estimation during training and arithmetic coding during inference.
To further improve low-bitrate efficiency, ECC introduces entropy skip, which omits highly predictable residual symbols using decoder-available scale estimates without transmitting additional skip masks.
Extensive experiments show that ECC achieves a favorable low-bitrate rate--distortion trade-off over conventional and neural codec baselines, reducing BD-rate by 39.9\% on ViSQOL and 76.3\% on PESQ on average over two widely-used test sets.
Ablation and diagnostic studies further validate the effectiveness of entropy modeling.
Project Page: \url{https://avery-xu.github.io/ECC-demo/}

\end{abstract}


\input{sections/introduction}

\input{sections/problem_formulation}

\input{sections/overview_of_progress}

\input{sections/entropy_modeling_motivation}

\input{sections/methodology}

\input{sections/experiment}

\input{sections/conclusion}



\bibliographystyle{IEEEtran}
\bibliography{reference}

\vspace{-10 mm} 
\begin{IEEEbiography}
[{\includegraphics[width=1in,height=1.25in,clip,keepaspectratio]{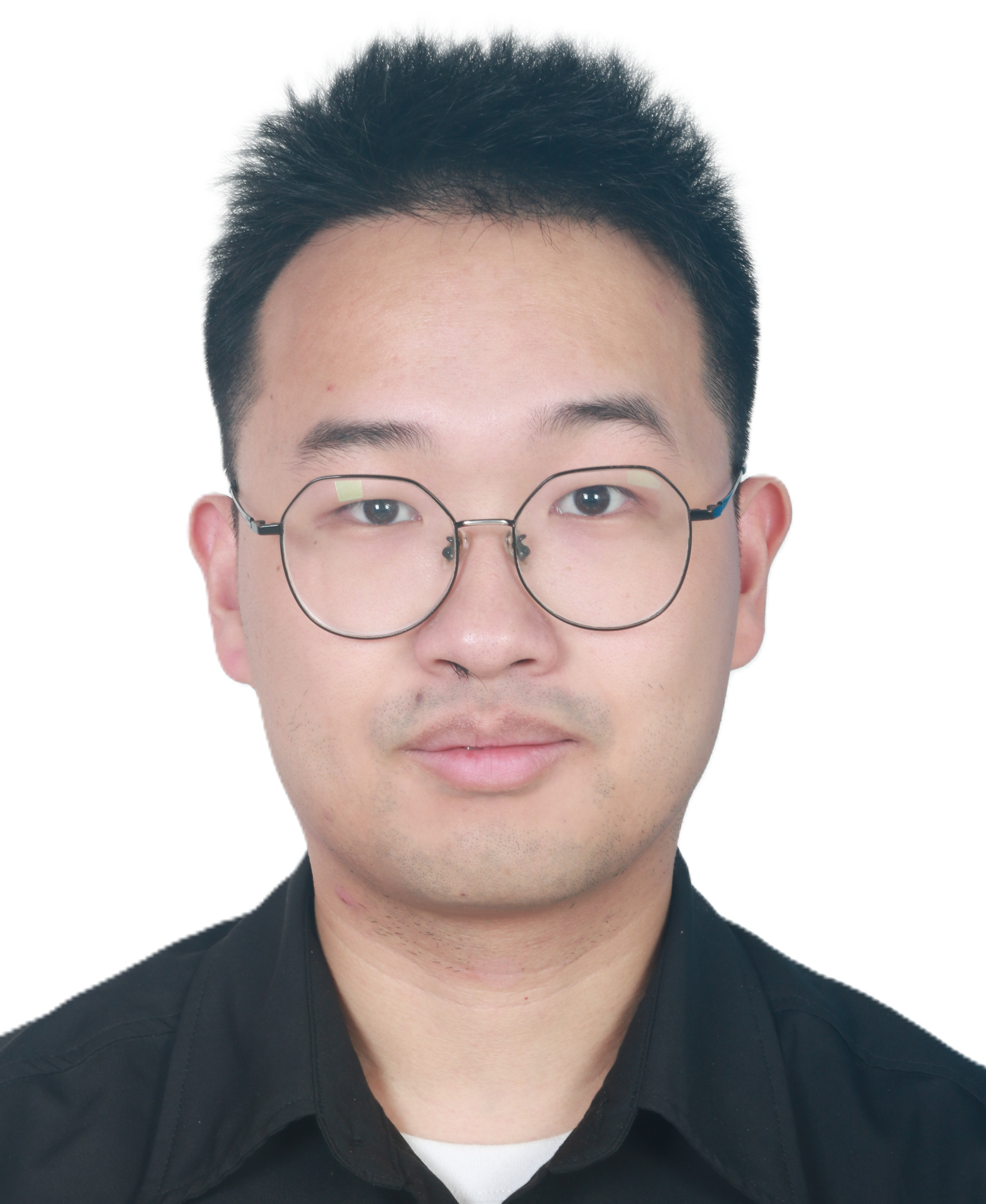}}]
{Jun Xu} receives the B.E. degree from Shanghai Jiao Tong University, Shanghai, China in 2020. He is currently pursuing the Ph.D degree with the Department of Electronic Engineering, Shanghai Jiao Tong University, Shanghai, China. His research interests include audio/video compression and multimedia system.
\end{IEEEbiography}
\vspace{-10 mm} 
\begin{IEEEbiography}
[{\includegraphics[width=1in,height=1.25in,clip,keepaspectratio]{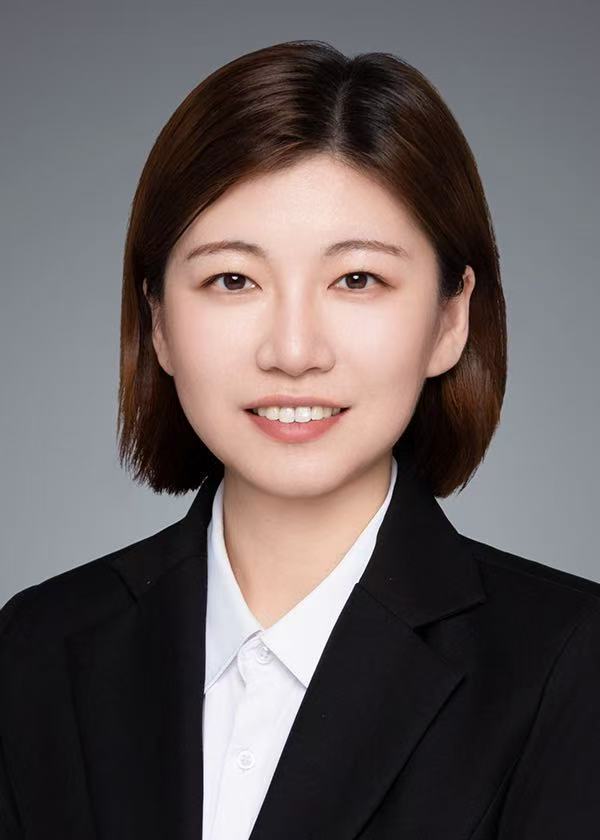}}]
{Zhengxue Cheng}
(Member, IEEE) receives the B.E. degree from Shanghai Jiao Tong University, Shanghai, China in 2014 and the M.E. degrees from Waseda University, Kitakyushu, Japan and Shanghai Jiao Tong University in 2015 and 2017, respectively through a double-degree program. She receives a PhD.degree at Waseda University, Tokyo, Japan in 2020. Then She worked in Ant Group, Hangzhou, China, as an Algorithm Expert until April 2024. She joined the institute of Image Communication and Network Engineering, Shanghai Jiao Tong University as an assistant researcher in May 2024. Her research interests include deep learning-based media compression and quality evaluation.
\end{IEEEbiography}
\vspace{-10 mm} 
\begin{IEEEbiography}
[{\includegraphics[width=1in,height=1.25in,clip,keepaspectratio]{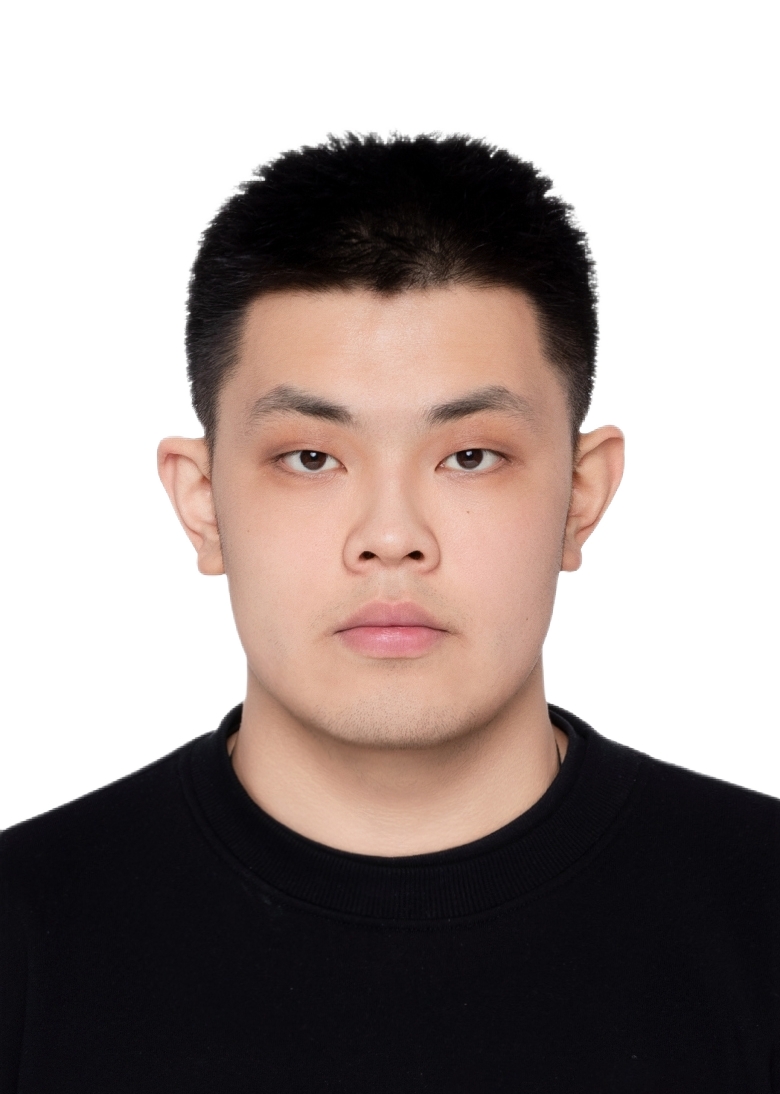}}]
{Fengxi Zhang} received the B.E. degree from Xidian University, Shaanxi, China in 2024. He is currently pursuing the Ph.D. degree with the Department of Electronic Engineering, Shanghai Jiao Tong University, Shanghai, China. His research interests include audio compression.
\end{IEEEbiography}
\vspace{-10 mm} 
\begin{IEEEbiography}
[{\includegraphics[width=1in,height=1.25in,clip,keepaspectratio]{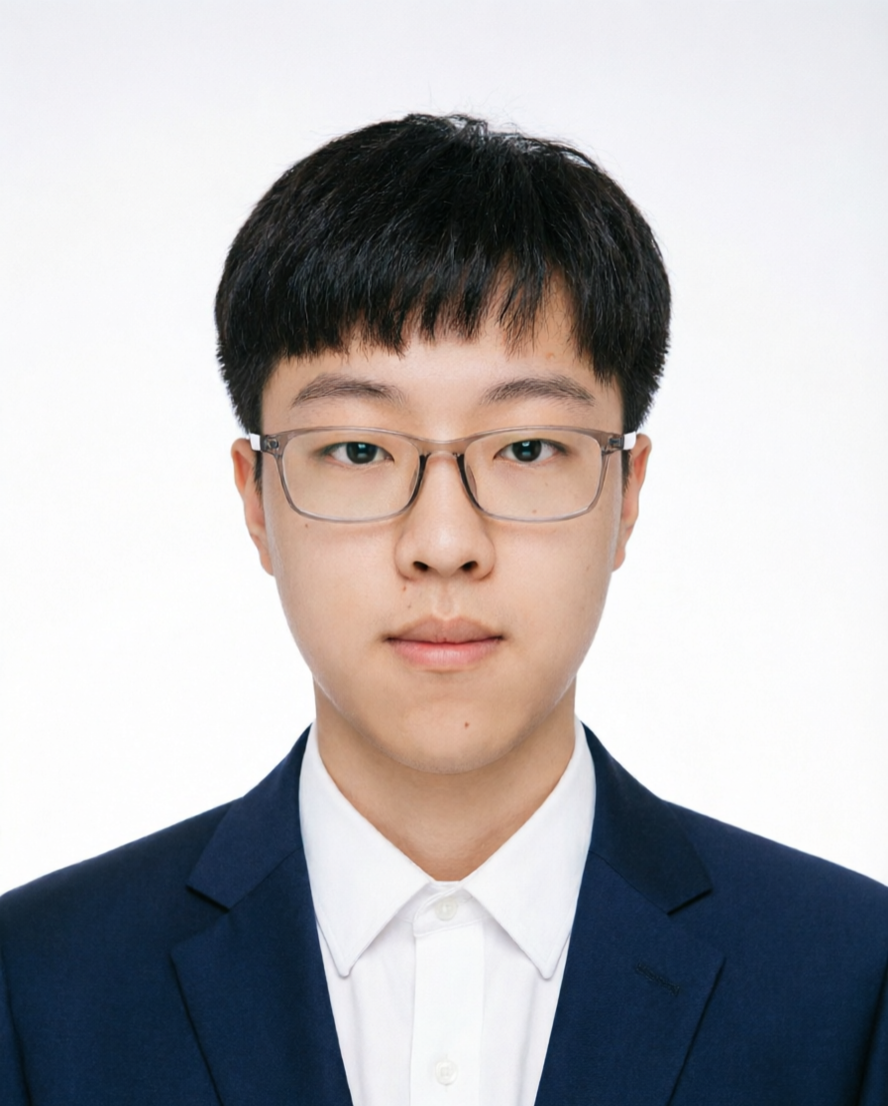}}]
{Yuhan Liu} received the B.S. degree from Shanghai Jiao Tong University, Shanghai, China. He is currently pursuing the M.S. degree with SJTU Paris Elite Institute of Technology, Shanghai Jiao Tong University. His research interests include audio compression and image compression.
\end{IEEEbiography}
\vspace{-10 mm} 
\begin{IEEEbiography}[{\includegraphics[width=1in,height=1.25in,clip,keepaspectratio]{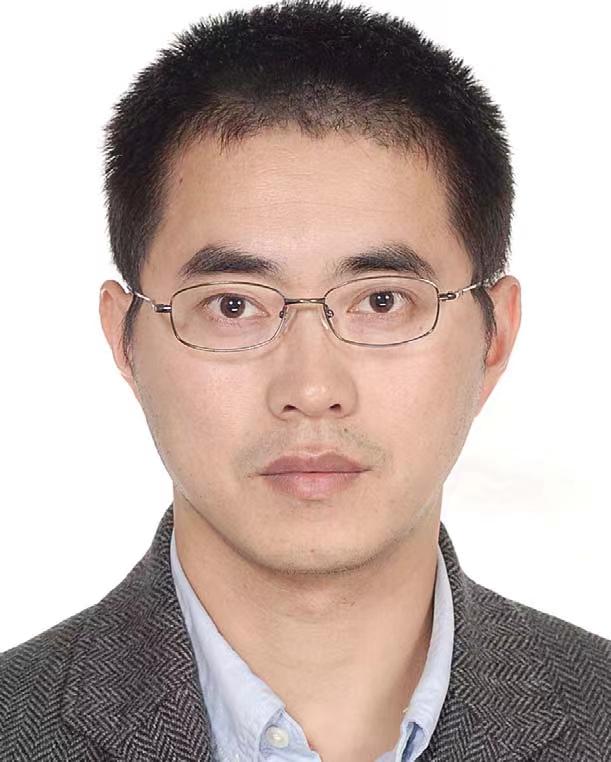}}]
{Li Song} (Senior Member, IEEE) received the B.E. and M.S. degrees in engineering in 1997 and 2000, respectively, and the Ph.D. degree in electrical engineering from Shanghai Jiao Tong University (SJTU) in 2005. He is currently a full professor with the department of electronic engineering. He was also a visiting professor with Santa Clara University from 2011 to 2012. He has 300 publications, 50 granted patents, and 20 standard technical contributions. His research interests include visual signal processing and artificial intelligent for multimedia. He has been serving as an associate editor for Multidimensional Systems and Signal Processing from 2012 to 2018 and anassociate editor for the IEEE Transactions on Broadcasting since 2024.
\end{IEEEbiography}
\vspace{-10 mm} 
\begin{IEEEbiography}[{\includegraphics[width=1in,height=1.25in,clip,keepaspectratio]{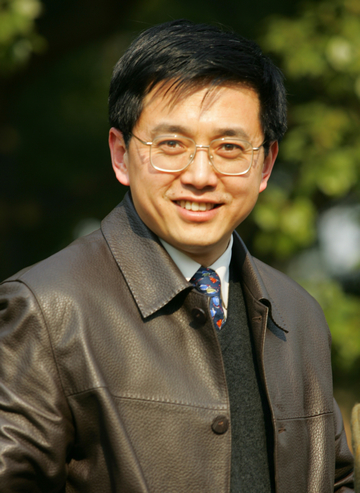}}]{Wenjun Zhang}
(Fellow, IEEE) received B.S., M.S., and Ph.D. degrees in electronic engineering from Shanghai Jiao Tong University, Shanghai, China, in 1984, 1987, and 1989, respectively.
From 1990 to 1993, he worked as a Postdoctoral Fellow with Philips, Nuremberg, Germany, where he was actively involved in developing the HD-MAC system.
He joined the faculty of Shanghai Jiao Tong University in 1993 and became a Full Professor of Electronic Engineering in 1995. He is the Chief Scientist of the Chinese Digital TV Engineering Research Centre, an industry/government consortium in DTV technology research and standardization.
His main research interests include digital video coding and transmission, multimedia semantic processing, and intelligent video surveillance.
\end{IEEEbiography}

\vfill

\end{document}

%% file: sections/introduction.tex
\section{Introduction}
\label{sec:intro}

\begin{figure}[t]
    \centering
    \includegraphics[trim=1mm 0mm 0mm 0mm, clip, width=1\linewidth]{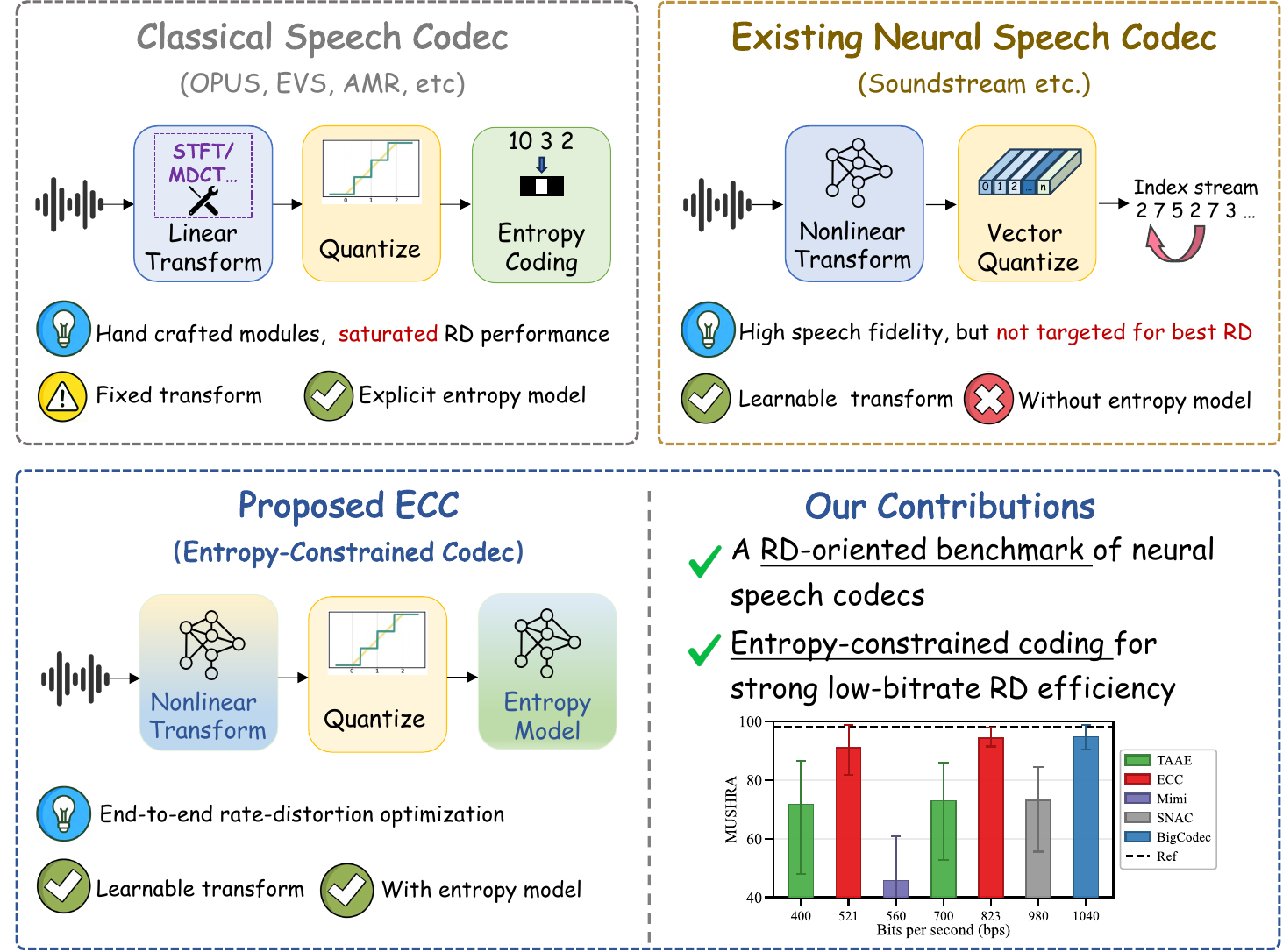}
    \vspace{-7mm}
    \caption{Positioning of Proposed ECC. Conventional codecs rely on hand-designed transforms, quantization rules, and entropy-coding tools. Recent neural speech codecs learn nonlinear representations but often describe their quantized latents using preset-rate indices or symbols, leaving probability modeling decoupled from representation learning. ECC integrates learned entropy modeling into the neural transform-coding pipeline, so that scalar latents are optimized for both reconstruction quality and statistical compressibility under an end-to-end rate--distortion objective.}
    \label{fig:teaser}
\vspace{-5mm}
\end{figure}

\IEEEPARstart{S}{peech} compression is essential for representing speech signals under constrained transmission, storage, and computational budgets.
It is particularly important for low-bitrate communication scenarios, such as mobile, real-time, and satellite speech services, where intelligibility and perceptual quality must be preserved with only a few hundred to a few thousand bits per second.
Recent 3GPP standardization efforts on non-terrestrial networks and ultra-low-bitrate speech codecs further highlight the need for efficient speech coding under extremely limited bit budgets~\cite{3gpp_ntn,3gpp_tr26940}.

Conventional codecs, such as AAC~\cite{aac}, Opus~\cite{opus}, EVS~\cite{evs}, and AMR~\cite{amr}, have long supported speech and audio communication through carefully engineered transform, prediction, quantization, and entropy-coding modules.
These pipelines are reliable and deployment-friendly, but their coding efficiency is ultimately constrained by hand-designed signal models and manually optimized coding tools.
As illustrated in Fig.~\ref{fig:teaser}, this handcrafted transform-coding paradigm differs fundamentally from recent neural codecs, which learn nonlinear speech representations from data.

Recent neural speech codecs have moved beyond hand-designed signal models by learning nonlinear analysis and synthesis transforms directly from data.
Following the transform-coding paradigm~\cite{TC}, most of them adopt an encoder-quantizer-decoder pipeline and improve low-bitrate quality through stronger backbones, perceptual losses, and discrete latent representations~\cite{soundstream,encodec,dac,funcodec,moshi}.
Despite these advances, many existing codecs still describe their quantized latents using preset-rate discrete symbols, whose nominal cost is determined by frame rate, quantizer depth, and codebook size.
Therefore, the learned representation and the actual coding distribution are often optimized in a decoupled manner.

From a source-coding perspective, this decoupling creates a mismatch between the generated symbol streams and the probability distributions used for coding.
Speech latents and codec indices are not uniformly distributed random symbols; instead, they inherit strong predictability from pitch periodicity, phonetic continuity, speaker-dependent dynamics, and temporal context.
Fixed-length coding assigns the same number of bits to frequent and rare symbols, and therefore cannot exploit the marginal non-uniformity and temporal dependency of the generated symbol streams.
Post-hoc entropy coding can reduce the lossless storage or transmission cost of a fixed symbol stream, but it cannot reshape the latent representation that produced the stream.
This motivates entropy-constrained training, where the transform, quantizer, and probability model are optimized jointly so that the learned latents are both reconstructive and statistically compressible.

In this paper, we benchmark neural speech compression from a rate--distortion perspective, with a particular focus on entropy-constrained coding.
We provide a unified formulation and benchmark-style analysis of recent neural speech codecs, revealing that explicit probability modeling remains underexplored.
We further propose ECC, an Entropy-Constrained Codec that integrates learned entropy modeling and entropy skip into scalar-latent speech coding, and validate its effectiveness through objective, subjective, ablation, and diagnostic experiments.

The main contributions are summarized as follows:
\begin{itemize}
    \item We present a unified formulation and RD-oriented benchmark analysis of recent neural speech codecs, clarifying the gap between preset-rate discrete representations and learned probability modeling.
    \item We propose \textbf{ECC}, a novel Entropy-Constrained Codec that integrates scalar quantization, channel-wise probabilistic entropy modeling, and entropy skip for highly predictable latents using end-to-end rate--distortion optimization for speech compression.
    \item We provide comprehensive objective, subjective, ablation, complexity, post-hoc entropy-coding, and generalization evaluations, showing consistent low-bitrate RD advantages over conventional and neural codec baselines, including 44.2\%/35.7\% ViSQOL and 69.4\%/83.3\% PESQ BD-rate reductions over FunCodec~\cite{funcodec} on LibriTTS~\cite{libritts}/VCTK~\cite{vctk} datasets.
\end{itemize}

One related preliminary publication is our conference paper~\cite{iscas_ecc}, which explored an early rate-aware learned speech compression framework.
Different from that, this paper focuses on benchmarking neural speech compression from a rate--distortion perspective, with a unified formulation, a taxonomy of recent neural speech codecs, the entropy skip mechanism, and expanded objective, subjective, complexity, post-hoc entropy-coding, ablation, and generalization analyses.

%% file: sections/problem_formulation.tex
\section{Problem Formulation}
\label{sec:formulation}

This section establishes a unified mathematical formulation for learning-based speech compression.
Although existing neural codecs differ in their choice of input domain, transform architecture, quantization mechanism, and entropy coding strategy, they inherently share a common data path from waveform representations to continuous latents, discrete symbols, bitstreams, and final reconstruction.

\subsection{Basic Pipeline of Learning-based Speech Coding}

Let $x \in \mathbb{R}^{T}$ denote an input speech waveform.
A learning-based codec first maps $x$ to a signal-domain representation $u$ through a front-end transform,
\begin{equation}
    u = \mathcal{T}(x),
\end{equation}
where $\mathcal{T}$ can be identity mapping for time-domain coding or a predefined time--frequency transform such as STFT or MDCT.

An analysis transform $f_{\theta}(\cdot)$ then converts $u$ into a continuous latent sequence,
\begin{equation}
    y = f_{\theta}(u), \qquad y \in \mathbb{R}^{T' \times D},
\end{equation}
where $T'$ is the number of latent frames and $D$ is the latent dimensionality.
The quantizer $Q(\cdot)$ maps $y$ to discrete symbols $s$, and the corresponding dequantization or embedding lookup produces reconstructed latents $\hat{y}$,
\begin{equation}
    s = Q(y), \qquad \hat{y} = D_{Q}(s).
\end{equation}
Here, $s$ may denote codebook indices in VQ/RVQ-based codecs or integer-valued coordinates in SQ-based codecs.

For transmission or storage, the discrete symbols are losslessly converted into a binary bitstream $b$ by an entropy coder,
\begin{equation}
    b = \mathcal{B}(s; p_{c}), \qquad R(s) \approx - \log_{2} p_{c}(s),
\end{equation}
where $p_{c}$ denotes the coding distribution and $R(s)$ is the expected code length.
Fixed-length index coding corresponds to a uniform and non-adaptive coding distribution determined by the symbol alphabet size.
In contrast, entropy-constrained coding estimates $p_{c}$ from latent statistics, allowing arithmetic or range coding to achieve an expected code length close to $-\log_{2}p_{c}(s)$.

Finally, a synthesis transform $g_{\phi}(\cdot)$ reconstructs the signal-domain representation, and the inverse front-end transform returns it to the waveform domain,
\begin{equation}
    \hat{u} = g_{\phi}(\hat{y}), \qquad \hat{x} = \mathcal{T}^{-1}(\hat{u}).
\end{equation}
This pipeline provides a common view of recent neural speech codecs, regardless of whether they differ in domain, backbone architecture, quantization strategy, or entropy-coding design.

\subsection{Challenges and Opportunities}

The central challenge of learning-based speech compression is to learn latents that are reconstructive, compact, and statistically compressible.
They should preserve perceptually important speech information while producing discrete symbols whose probability structure can be efficiently modeled.

Speech signals contain temporal and spectral regularities, such as pitch periodicity, phonetic continuity, and speaker-dependent dynamics, which may remain as dependencies across time, channels, and quantization stages.
Preset-rate discrete representations ignore these statistics because their nominal cost is fixed by the quantizer configuration, while post-hoc entropy coding can only compress an already generated symbol stream.
This motivates entropy-constrained learning, where the probability model provides a differentiable rate term during training and coding probabilities during inference, enabling the transform, quantizer, and entropy model to jointly produce latents that are easier to entropy-code.

%% file: sections/overview_of_progress.tex
\section{Benchmarking Recent Neural Speech Codecs}
\label{sec:progress}
\begin{figure*}[t]
    \centering
    \includegraphics[width=1 \textwidth]{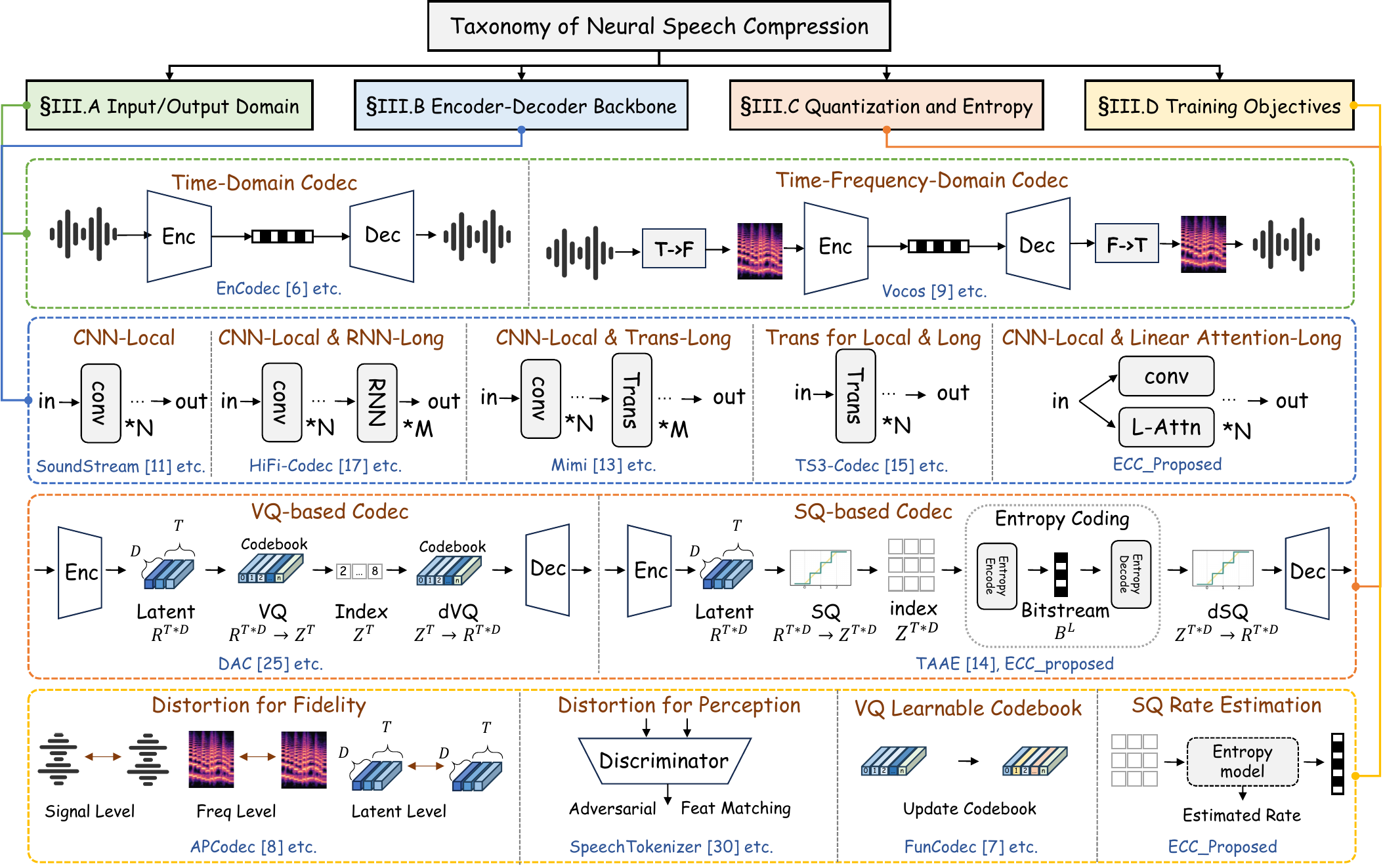}
    \vspace{-8mm}
    \caption{Taxonomy of recent learning-based speech compression methods.
    We organize the design space along four axes: input/output domain, encoder--decoder backbone, quantization and entropy coding, and training objectives.
    The taxonomy highlights how codec designs differ in signal representation, temporal modeling, discrete symbol construction, and whether probability modeling is integrated into training and coding.}
    \label{fig:taxonomy}
\vspace{-5mm}
\end{figure*}

\begin{table*}[htbp]
\centering
\caption{Summary of Recent Neural Speech Codec Works}
\label{tab:codec_summary}
\setlength{\tabcolsep}{6pt}
\begin{tabular}{llllllll}
\toprule
\textbf{Name} & \textbf{Venue/Year} & \textbf{Domain} & \textbf{Encoder} & \textbf{Decoder} & \textbf{Quantizer} & \textbf{Entropy} & \textbf{Training Objective} \\
\midrule
SoundStream \cite{soundstream} & TASLP 2021 & Time & CNN & CNN & RVQ & \ding{55} & GAN, Feat, Rec\\
EnCodec\cite{encodec} & TMLR 2023 & Time & CNN+RNN & CNN+RNN & RVQ & \ding{55} & GAN, Feat, Rec, VQ\\
DAC\cite{dac} & NeurIPS 2023 & Time & CNN & CNN & RVQ & \ding{55}& GAN, Feat, Rec, VQ\\
HiFi-Codec\cite{hifi-codec} & arXiv 2023 & Time & CNN+RNN & CNN+RNN & GRVQ & \ding{55} & GAN, Feat, Rec, VQ\\
AudioDec\cite{audiodec} & ICASSP 2023 & Time & CNN & CNN & RVQ & \ding{55} & GAN, Feat, Rec, VQ\\
ESC\cite{esc} & arXiv 2024 & Time+Freq & Trans & Trans & CSRVQ &\ding{55} & Rec, VQ\\
Vocos\cite{vocos} & ICLR 2024 & Time+Freq & CNN & CNN & RVQ &\ding{55}& GAN, Feat, Rec\\
NDVQ\cite{ndvq} & SLT 2024 & Time & CNN+RNN & CNN+RNN & RNDVQ &\ding{55}& GAN, Feat, Rec, VQ\\
SNAC\cite{snac} & NeurIPS WS 2024 & Time & CNN+RNN & CNN & MSRVQ &\ding{55} & GAN, Feat, Rec, VQ\\
FunCodec\cite{funcodec} & ICASSP 2024 & Time+Freq & CNN+RNN & CNN+RNN & RVQ & \ding{55} & GAN, Feat, Rec, VQ\\
SpeechTokenizer\cite{speechtokenizer} & ICLR 2024 & Time & CNN+RNN & CNN & RVQ & \ding{55} & GAN, Feat, Rec, VQ\\
APCodec\cite{apcodec} & TASLP 2024 & Time+Freq & CNN & CNN & RVQ &\ding{55} & GAN, Feat, Rec, VQ\\
MDCTCodec\cite{mdctcodec} & SLT 2024 & Time+Freq & CNN & CNN & RVQ &\ding{55} & GAN, Feat, Rec, VQ\\
Mimi\cite{moshi} & arXiv 2024 & Time & CNN+Trans & CNN+Trans & RVQ &\ding{55} & GAN, Feat, Rec, VQ\\
BigCodec\cite{bigcodec} & arXiv 2024 & Time & CNN+RNN & CNN+RNN & SVQ &\ding{55} & GAN, Feat, Rec, VQ\\
Spectral Codecs\cite{spectral} & arXiv 2024 & Time+Freq & CNN & CNN & FSQ & \ding{55} & GAN, Feat, Rec\\
SemanticCodec\cite{semanticodec} & JSTSP 2024 & Time & CNN+Trans & Trans & RVQ & \ding{55} & Diff, VQ\\
SQ-Codec\cite{sq-codec} & TASLP 2025 & Time & CNN & CNN & FSQ &\ding{55}& GAN, Rec\\
StreamCodec\cite{streamcodec} & SPL 2025 & Time+Freq & CNN & CNN & RSVQ & \ding{55} & GAN, Feat, Rec, VQ\\
TAAE\cite{taae} & ICLR 2025 & Time & CNN+Trans & CNN+Trans & FSQ &\ding{55}& GAN, Feat, Rec\\
TS3-Codec\cite{ts3codec} & INTERSPEECH 2025 & Time & Trans & Trans & SVQ & \ding{55}& GAN, Feat, Rec, VQ\\
WavTokenizer\cite{wavtokenizer} & ICLR 2025 & Time & CNN+Trans & CNN+Trans & SVQ &\ding{55} & GAN, Feat, Rec, VQ\\
FocalCodec\cite{focal} & NeurIPS 2025 & Time & CNN+Trans & CNN & BSQ &\ding{55} & GAN, Feat, Rec, Ent\\
SpecTokenizer\cite{spectokenizer} & INTERSPEECH 2025 & Time+Freq & CNN+RNN & CNN+RNN & SVQ &\ding{55} & GAN, Feat, Rec, Cmt\\
\textbf{ECC (Ours)} & This work & Time+Freq & CNN+L-Attn & CNN+L-Attn & SQ & \ding{51} & GAN, Feat, Rec, Rate\\
\bottomrule
\end{tabular}

\vspace{1mm}
\begin{minipage}{0.98\textwidth}
\scriptsize
\emph{Note:} The ``Entropy'' column indicates whether an explicit learned probability model is integrated into codec training or latent coding; post-hoc lossless compression of generated indices is discussed separately.
Objective abbreviations: Rec, reconstruction loss; GAN, adversarial loss; Feat, feature matching; VQ, vector-quantization loss; Cmt, commitment loss; Ent, entropy-related auxiliary loss; Diff, diffusion objective; MP, masked prediction; Rate, explicit rate term.
\end{minipage}
\vspace{-5mm}
\end{table*}

This section reviews recent learning-based speech codecs under the unified pipeline in Section~\ref{sec:formulation}.
As summarized in Fig.~\ref{fig:taxonomy} and Table~\ref{tab:codec_summary}, existing methods are organized along four axes: input/output domain, encoder--decoder backbone, quantization and entropy coding, and training objective.
This taxonomy highlights how different codec designs represent speech signals, model temporal dependencies, construct discrete symbols, and handle coding costs.

\subsection{Input and Output Domains}

Existing codecs differ first in the signal domain where neural coding is performed.
Time-domain codecs directly process waveforms, as in SoundStream~\cite{soundstream} and EnCodec~\cite{encodec}, keeping the signal path simple while requiring the neural transform to learn both local waveform patterns and spectral structure.
Time--frequency-domain codecs introduce explicit spectral front ends, such as STFT features in FunCodec~\cite{funcodec}, amplitude--phase modeling in APCodec~\cite{apcodec}, inverse-transform-aware decoding in Vocos~\cite{vocos}, and MDCT preprocessing in MDCTCodec~\cite{mdctcodec}.
These designs expose spectral sparsity and harmonic regularities before coding, but also introduce choices on windowing, hop size, phase representation, and inverse transform design.
Overall, time-domain coding favors end-to-end waveform modeling, whereas time--frequency-domain coding injects signal-structure priors that may ease representation learning and latent compression.

\subsection{Encoder-Decoder Backbone}

The encoder--decoder backbone determines how local acoustic details and long-range speech dependencies are represented before quantization.
CNN-based codecs, such as SoundStream~\cite{soundstream} and DAC~\cite{dac}, are efficient and deployment-friendly, but their finite receptive fields can limit long-context modeling.
Hybrid CNN--RNN systems, including EnCodec~\cite{encodec} and HiFi-Codec~\cite{hifi-codec}, add recurrent temporal modeling to convolutional features.
Recent codecs further introduce attention-based modules, such as the CNN--Transformer design in Mimi~\cite{moshi} and Transformer-heavy structures in TAAE~\cite{taae} and TS3-Codec~\cite{ts3codec}.
These designs improve temporal context modeling, but usually require more computation, data, and model capacity.
This trend suggests that effective speech coding benefits from combining efficient local modeling with lightweight long-range dependency modeling.

\subsection{Quantization and Entropy}

Quantization maps continuous latents into discrete symbols for compression.
VQ-based designs remain dominant, with RVQ widely used since SoundStream~\cite{soundstream} because successive codebooks progressively refine residual errors.
Recent variants improve capacity, utilization, or robustness through grouped quantization~\cite{hifi-codec}, probabilistic residual selection~\cite{ndvq}, multi-resolution or cross-scale quantization~\cite{snac,esc}, factorized codebooks~\cite{dac}, and streamable scalar-vector designs~\cite{streamcodec}.
Simpler quantizers have also gained traction.
SVQ-based systems, including BigCodec~\cite{bigcodec}, TS3-Codec~\cite{ts3codec}, and WavTokenizer~\cite{wavtokenizer}, reduce quantization depth by relying on stronger transforms.
FSQ and related scalar designs~\cite{fsq} are adopted in SQ-Codec~\cite{sq-codec}, Spectral Codecs~\cite{spectral}, and FocalCodec~\cite{focal}, reducing codebook management complexity.

From a compression perspective, many neural speech codecs still rely on nominal token rates determined by codebook size, quantizer count, and frame rate.
Post-hoc entropy coding, as used for EnCodec-style RVQ indices~\cite{encodec}, reduces the lossless index-stream cost but does not affect the learned representation.
Entropy-constrained coding instead integrates probability modeling and rate estimation into training, making coding cost part of representation learning.
This principle has been extensively studied in learned image compression, from factorized priors, content-weighted priors, hyperpriors, and conditional probability models~\cite{balle2017,li2021content,mentzer2018conditional,balle2018}, to autoregressive--hierarchical models and stronger likelihood models~\cite{minnen2018,cheng20,bross2021vvc}.
Later work further improves the accuracy--latency trade-off with masked, checkerboard, channel-wise, space--channel, Transformer, hierarchical, and dictionary-based contexts~\cite{pixelcnnpp,checkerboard,minnen2020,elic,m2t,maskgit,vit,entroformer,contextformer,mlicpp,groupedmixer,hu2022benchmark,qarv,dbem}.
For speech compression, the same principle suggests modeling side information, decoded channel context, and temporal context during training, rather than treating entropy coding as a post-hoc stage.

\subsection{Training Objectives}

As summarized in Table~\ref{tab:codec_summary}, training objectives usually combine reconstruction fidelity, perceptual quality, quantization stability, and coding-related constraints.
Reconstruction losses (\textbf{Rec}) are computed in waveform, spectral/mel, or latent spaces, while VQ-based codecs add codebook or commitment losses (\textbf{VQ}/\textbf{Cmt}) to stabilize discrete representation learning.
Related commitment terms are also used in scalar or lookup-free tokenizers such as SpecTokenizer~\cite{spectokenizer}.
Perceptual quality is commonly improved with adversarial losses (\textbf{GAN}) and feature matching (\textbf{Feat}), and diffusion objectives (\textbf{Diff}) are used in generative decoders such as SemanticCodec~\cite{semanticodec}.

Coding-aware objectives remain less common.
FocalCodec~\cite{focal} introduces an entropy-related auxiliary term (\textbf{Ent}), whereas the proposed codec uses an explicit learned rate term (\textbf{Rate}) from the probability model; Section~\ref{sec:motivation} gives the RD formulation.
Beyond rate--distortion optimization, other objectives shape token semantics through masked prediction~\cite{bert,hubert,wavlm,gslm,mousavi2024semantic}, source or speaker disentanglement~\cite{yang2021sourceaware,omran2023disentangling,polyak2021resynthesis,jiang2023disentangled,ticodec,facodec,lscodec,socodec,ecapa,sdcodec}, SSL or LLM distillation~\cite{speechtokenizer,xcodec,moshi,llmcodec,llama2,semanticodec}, and supervised phonetic or ASR-style tokenization~\cite{cosyvoice,cosyvoice2,past,otcfm,glm4voice}.

%% file: sections/entropy_modeling_motivation.tex
\begin{figure*}[t]
    \centering
    \includegraphics[trim=0mm 0mm 1mm 0mm, clip, width=1\textwidth]{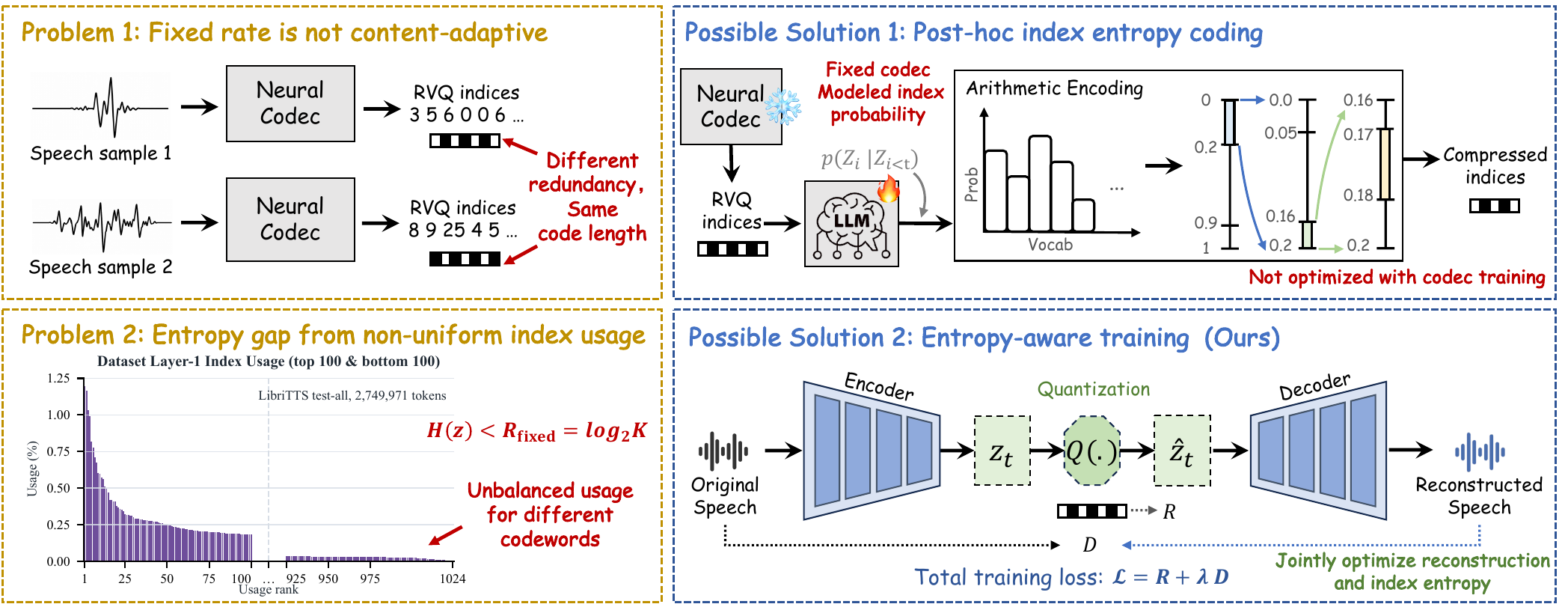}
    \vspace{-7mm}
    \caption{Motivation for entropy-aware neural speech coding.
    The left part illustrates two sources of redundancy in fixed-length RVQ indices: content-independent rate allocation and non-uniform codeword usage.
    Speech contents with the same duration are assigned the same nominal code length under a preset RVQ configuration, although their redundancy can differ substantially; moreover, dataset-level codeword usage is highly non-uniform, so the index entropy $H(Z)$ can be lower than the fixed-length rate $R_{\mathrm{fixed}}$.
    The right part contrasts post-hoc index entropy coding with the proposed entropy-aware training, where the rate estimate participates in representation learning.}
    \label{fig:rvq_redundancy}
\vspace{-5mm}
\end{figure*}

\section{Motivation for Entropy Modeling}
\label{sec:motivation}

Fig.~\ref{fig:rvq_redundancy} motivates entropy-aware coding for neural speech codec symbols.
Fixed-rate RVQ remains attractive for deployment because it provides simple rate control and stable token budgets.
However, from a source-coding perspective, preset-rate symbol streams can be suboptimal when their distributions are non-uniform and temporally dependent.
We first identify two inefficiencies in RVQ index streams, and then contrast post-hoc index coding with end-to-end entropy-aware training.

\subsection{Two Inefficiencies of RVQ Indices}

Many VQ/RVQ-based neural speech codecs use a preset token rate.
For an RVQ module with $N_q$ stages and codebook size $K_j$ at stage $j$, the fixed-length rate per latent frame is
\begin{equation}
    R_{\mathrm{fixed}} = \sum_{j=1}^{N_q} \log_2 K_j.
    \label{eq:rvq_rate}
\end{equation}
At a fixed frame rate, an utterance with $T'$ latent frames therefore costs $T'R_{\mathrm{fixed}}$ bits regardless of its content.
This is the first limitation shown in Fig.~\ref{fig:rvq_redundancy}: for the same frame count, an information-rich speech segment and a highly predictable or repetitive segment are assigned the same number of RVQ index bits.
The resulting index cost is fixed before observing the actual symbol statistics, and therefore cannot reflect the different predictability of different speech segments.

The second limitation is uneven codeword usage.
Fixed-length index coding implicitly treats all entries in a codebook as equally costly, but trained RVQ codebooks are often used non-uniformly.
For an utterance, let $\mathbf{Z}=\{Z_{t,j}:1\le t\le T',\,1\le j\le N_q\}$ be the full RVQ index sequence.
An ideal entropy coder approaches the sequence entropy $H(\mathbf{Z})$, while fixed-length coding spends $T'R_{\mathrm{fixed}}$ bits.
Under an empirical distribution of RVQ indices, the ideal redundancy relative to fixed-length coding can be decomposed as
\begin{align}
    \Delta R
    &= T'R_{\mathrm{fixed}} - H(\mathbf{Z})
    = \Delta R_{\mathrm{marg}}+\Delta R_{\mathrm{dep}},\\
    \label{eq:rate_redundancy}
    \Delta R_{\mathrm{marg}}
    &= \sum_{t,j}\left[\log_2 K_j - H(Z_{t,j})\right], \\
    \Delta R_{\mathrm{dep}}
    &= \sum_{t,j}H(Z_{t,j}) - H(\mathbf{Z}).
\end{align}
The marginal $\Delta R_{\mathrm{marg}}$ measures loss caused by non-uniform codeword usage: if a codebook entry distribution is skewed, then $H(Z_{t,j})<\log_2 K_j$ and fixed-length coding wastes bits.
The dependency term $\Delta R_{\mathrm{dep}}$ captures additional predictability across time and quantization stages.
Thus, even before changing the neural codec itself, generated RVQ indices contain exploitable statistical structure beyond fixed-length coding.

\subsection{Solution 1: Post-Hoc Index Coding}

A direct response is to keep the pretrained RVQ codec unchanged and add an index coder after training.
Given a fixed index sequence $\mathbf{i}$, a post-hoc coder estimates a marginal or conditional distribution, such as $q_{\eta}(i_t\mid i_{<t})$, and passes the resulting probabilities to an arithmetic coder.
Its expected index-stream length is
\begin{equation}
    R_{\mathrm{index}}
    \approx
    \mathbb{E}\big[-\log_2 q_{\eta}(\mathbf{i})\big].
\end{equation}
This strategy reduces bitrate by using arithmetic coding probabilities for frequent or predictable indices, while remaining compatible with existing codecs because encoder, RVQ codebooks, and decoder are unchanged.
However, it is only a lossless compression stage over a fixed representation: it can reduce transmitted index cost at a given reconstruction point, but cannot improve reconstruction quality, codebook usage, or the transform that generated the indices.
Therefore, the representation is not trained to become easier to entropy-code.

\subsection{Solution 2: Entropy-Aware Training}

Our codec uses the second solution in Fig.~\ref{fig:rvq_redundancy}: the rate model is included in representation learning itself.
Instead of generating fixed-cost RVQ indices and compressing them afterward, it estimates conditional probabilities for quantized scalar latents from side information and decoded context:
\begin{equation}
    p(\hat{\mathbf{y}} \mid \psi)
    = \prod_{t=1}^{T'} \prod_{c=1}^{D}
    p(\hat{y}_{t,c} \mid \mathcal{C}_{t,c}, \psi),
    \label{eq:entropy_model}
\end{equation}
where $\psi$ denotes side information and $\mathcal{C}_{t,c}$ is the available temporal or channel-wise context for element $(t,c)$.
This probability model provides a differentiable rate estimate,
\begin{equation}
    R_{\mathrm{latent}} \approx \mathbb{E}\left[-\log_{2} q(\hat{y}\mid \psi, C)\right],
\end{equation}
which can be optimized jointly with reconstruction losses.
The rate term includes side information and the conditional cost of the primary latents:
\begin{equation}
    R_{\mathrm{latent}}
    =
    -\log_2 p(\hat{\mathbf{z}})
    -\log_2 p(\hat{\mathbf{y}} \mid \hat{\mathbf{z}}, \mathcal{C}),
\end{equation}
where $\hat{\mathbf{z}}$ is transmitted side information and $\mathcal{C}$ is decoded context.
For fixed-rate RVQ, $R$ is a nominal token rate; for post-hoc index coding, $R$ is the compressed length of a fixed index stream; for entropy-aware training, $R$ is a learned entropy estimate jointly optimized with the transform, quantizer, entropy model, and decoder.
Scalar quantization avoids learned codebook lookup and codebook-utilization balancing, while the entropy model captures the marginal and conditional structure needed for efficient compression.

%% file: sections/methodology.tex
\section{Methodology}
\label{sec:method}

\begin{figure*}[t]
    \centering
    \includegraphics[trim=0mm 0mm 7mm 0mm, clip, width=1 \textwidth]{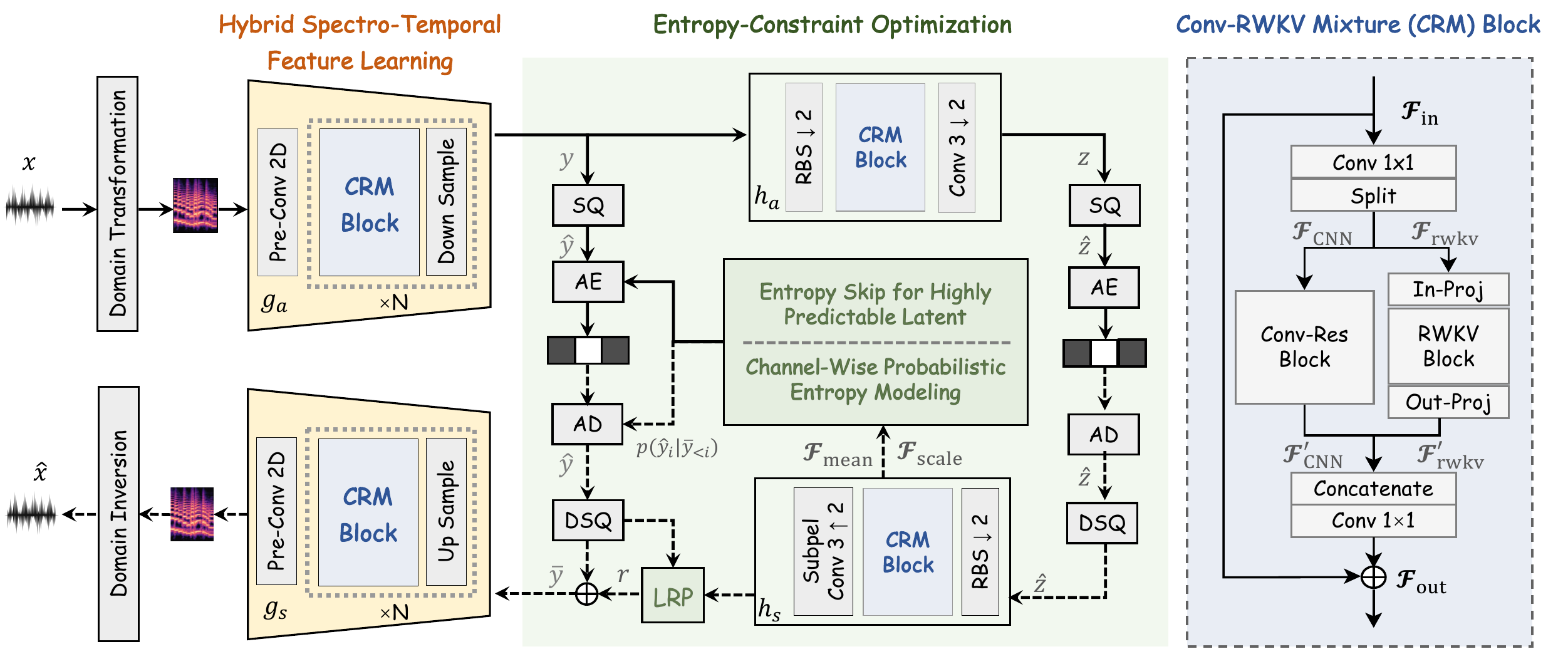}
    \vspace{-7mm}
    \caption{Overview of the proposed Entropy-Constrained Codec (ECC).
    ECC uses STFT-domain analysis--synthesis transforms with CRM blocks, scalar quantization, a hyperprior, and a channel-wise entropy model for latent probability estimation.
    Latent residual prediction (LRP) refines decoded slices, while entropy skip omits highly predictable residual symbols according to decoder-available scale estimates.
    Skipped residuals are reconstructed as zeros, and only non-skipped symbols are arithmetic coded.
    Here, SQ/DSQ denote scalar quantization/dequantization, and AE/AD denote arithmetic encoding/decoding.}
    \label{fig:framework}
\vspace{-5mm}
\end{figure*}

\subsection{Overview of the Proposed Framework}

Following Section~\ref{sec:motivation}, ECC jointly learns the analysis--synthesis transform, the latent probability model, and the reconstruction objective.
As shown in Fig.~\ref{fig:framework}, the waveform $x\in\mathbb{R}^{T}$ is first converted to the STFT representation $X_{\mathrm{stft}}\in\mathbb{C}^{F\times T'}$ and mapped by the analysis transform $g_a$ to the primary latent $y$.
A hyper-analysis transform $h_a$ further generates the hyper-latent $z$, whose quantized version $\hat{z}$ is transmitted as side information and decoded by $h_s$ to provide context features.

The primary latent is coded slice by slice using a channel-wise entropy model.
For each slice, hyperprior features and previously decoded slices predict the probability distribution used for rate estimation during training and arithmetic coding during inference.
Entropy skip omits highly predictable residual symbols, LRP refines decoded slices, and the synthesis transform followed by iSTFT reconstructs the waveform.
By combining channel-wise context with lightweight temporal modeling, ECC exploits both inter-channel dependency and long-range speech structure for probability estimation.

\subsection{Spectro-Temporal Analysis and Synthesis Transform}
\label{subsec:backbone}

The transform operates in the time--frequency domain to exploit speech spectral sparsity and locality.
The encoder first converts the waveform into an STFT representation and applies convolutional downsampling with CRM blocks.
The decoder mirrors this hierarchy with transposed convolutions and finally reconstructs the waveform through iSTFT.
This analysis--synthesis path can be written as \begin{equation} \begin{aligned} X_{\mathrm{stft}} &= \mathrm{STFT}(x),\\ y &= g_a(X_{\mathrm{stft}};\phi),\\ \hat{x} &= \mathrm{iSTFT}(g_s(\bar{y};\theta)), \end{aligned} \end{equation} where $\bar{y}$ is the refined latent representation after quantization, entropy modeling, and LRP.

Each CRM block splits the input feature $\mathcal{F}_{\mathrm{in}}$ into two channel groups, $(\mathcal{F}_{\mathrm{cnn}},\mathcal{F}_{\mathrm{rwkv}})=\mathrm{Split}(\mathcal{F}_{\mathrm{in}})$.
The CNN branch captures local time--frequency patterns, while the RWKV branch provides linear-time long-range temporal modeling~\cite{seanet,rwkv,rwkv6}.
The two branches are fused by a $1\times1$ convolution and added back residually:
\begin{equation}
\small
    \mathcal{F}_{\mathrm{out}}
    = \mathcal{F}_{\mathrm{in}}
    + \mathcal{W}_{\mathrm{fuse}}\big(
    \mathrm{Concat}(
    \mathcal{H}_{\mathrm{CNN}}(\mathcal{F}_{\mathrm{cnn}}),
    \mathcal{H}_{\mathrm{RWKV}}(\mathcal{F}_{\mathrm{rwkv}})
    )\big),
\end{equation}
where $\mathcal{H}_{\mathrm{CNN}}$ and $\mathcal{H}_{\mathrm{RWKV}}$ denote the two branch transformations.
Across scales, shallow high-resolution stages use fewer RWKV layers, while deeper low-resolution stages use more layers to capture longer temporal dependencies at lower cost.

\begin{figure*}[t]
    \centering
    \includegraphics[trim=2mm 0mm 0mm 0mm, clip, width=1 \textwidth]{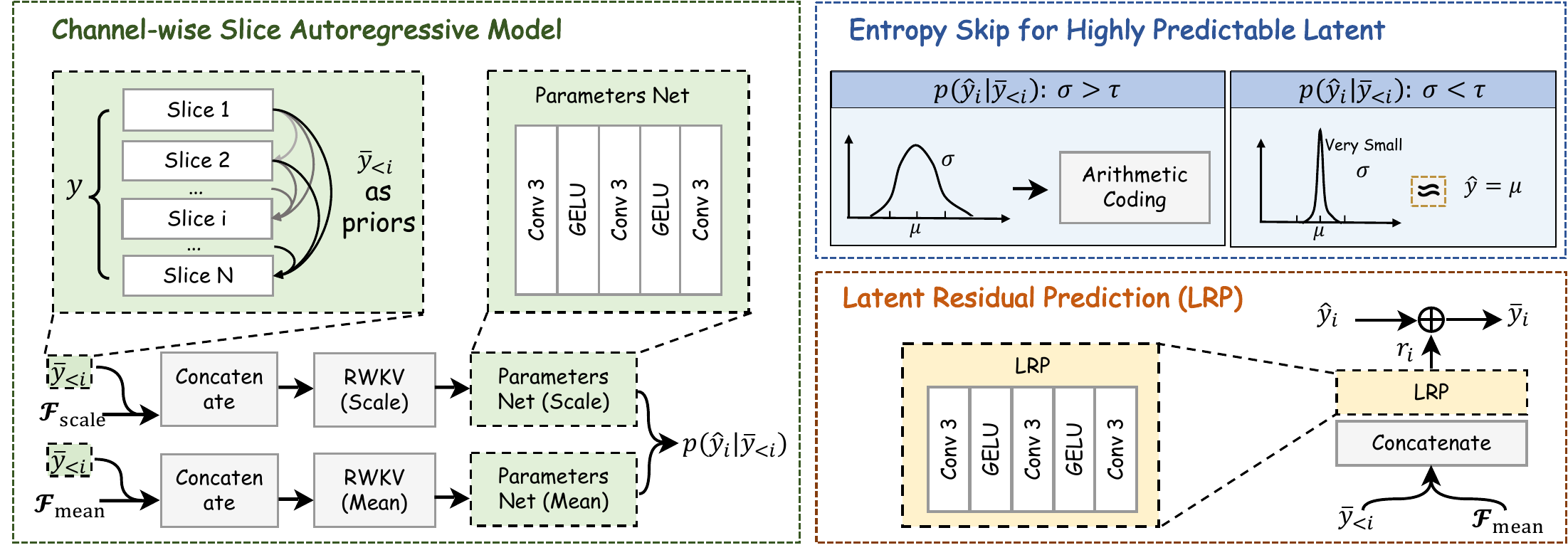}
    \vspace{-8mm}
    \caption{Channel-wise entropy model.
    The hyperprior path converts the primary latent $y$ into side information $\hat{z}$ and decodes it into context features $\mathcal{F}_{\mathrm{mean}}$ and $\mathcal{F}_{\mathrm{scale}}$.
    The channel-wise context model processes latent slices sequentially and predicts Gaussian parameters for each slice from the hyperprior features and previously decoded slices.
    LRP refines the decoded slice before it is passed to the synthesis transform and to subsequent context prediction.}
    \label{fig:entropy_model}
\vspace{-5mm}
\end{figure*}

\subsection{Channel-Wise Probabilistic Entropy Modeling}
\label{subsec:entropy}

The entropy model estimates scalar-latent code lengths and supplies decoder context.
As illustrated in Fig.~\ref{fig:entropy_model}, it consists of a hyperprior path, a channel-wise context model, and LRP.
The hyperprior path produces side information as
\begin{align}
    z &= h_a(y;\phi_h)\\
    \hat{z}&=Q(z),\\
    \mathcal{F}_{\mathrm{mean}},\mathcal{F}_{\mathrm{scale}}
    &= h_s(\hat{z};\theta_h).
\end{align}
The quantized hyper-latent $\hat{z}$ is coded with a factorized prior,
\begin{equation}
    p_{\hat{z}}(\hat{z})=\prod_j p_{\hat{z}_j}(\hat{z}_j),
\end{equation}
and is decoded before $y$, making $\mathcal{F}_{\mathrm{mean}}$ and $\mathcal{F}_{\mathrm{scale}}$ available to both encoder and decoder.
The transforms $h_a$ and $h_s$ also use CRM blocks to capture temporal structure.

The primary latent $y\in\mathbb{R}^{T' \times C_y}$ is evenly partitioned into $S$ channel slices $\{y_0,y_1,\dots,y_{S-1}\}$.
When coding slice $i$, the model conditions on the hyperprior features and previously refined slices $\bar{y}_{<i}$, and predicts scale and mean by
\begin{align}
    \sigma_i &= \mathcal{G}_{\mathrm{scale}}(\mathrm{Concat}(\mathcal{F}_{\mathrm{scale}}, \bar{y}_{<i})), \\
    \mu_i    &= \mathcal{G}_{\mathrm{mean}}(\mathrm{Concat}(\mathcal{F}_{\mathrm{mean}}, \bar{y}_{<i})).
\end{align}
The quantized-slice probability is modeled by a Gaussian density convolved with a unit-width uniform distribution:
\begin{equation}
    p_{\hat{y}_i|\hat{z},\bar{y}_{<i}}(\hat{y}_i)
    =
    \left(\mathcal{N}(\mu_i,\sigma_i^2)*
    \mathcal{U}\!\left(-\tfrac{1}{2},\tfrac{1}{2}\right)\right)(\hat{y}_i).
\end{equation}
Training uses uniform noise for differentiable likelihood estimation and deterministic rounding with a straight-through estimator on the reconstruction path.
At inference time, arithmetic coding uses the corresponding discrete probability mass.

After decoding, LRP compensates for scalar-quantization error before the slice is used by the synthesis transform and later context prediction.
Let $\tilde{y}_i$ denote the decoded symbol after the entropy-skip decision; without skip, $\tilde{y}_i=\hat{y}_i$.
The residual and refined slice are computed as
\begin{equation}
    r_i = \mathcal{G}_{\mathrm{LRP}}(\mathrm{Concat}(\tilde{y}_i, \mathcal{F}_{\mathrm{mean}}, \bar{y}_{<i})),
\end{equation}
\begin{equation}
    \bar{y}_i = \tilde{y}_i + r_i.
\end{equation}
The refined slice $\bar{y}_i$ is the latent representation consumed by both the decoder and subsequent entropy contexts.
Since LRP uses only the decoded slice, hyperprior features, and previously refined slices, it is fully decoder-available and does not introduce additional side information.

\subsection{Entropy Skip for Highly Predictable Latents}
\label{sec:entropy_skip}

Each primary-latent scalar is modeled by a Gaussian distribution conditioned on the hyperprior and already decoded context~\cite{balle2018,minnen2018}.
For element \(n\) in coding order, with predicted mean and scale \((\mu_n,\sigma_n)\), mean-centered scalar quantization codes the residual
\begin{equation}
    d_n = y_n-\mu_n,\quad
    \hat{d}_n=\mathrm{round}(d_n),\quad
    \hat{y}_n=\mu_n+\hat{d}_n .
    \label{eq:mean_centered_quant}
\end{equation}
Equivalently, arithmetic coding is applied to the integer residual symbol \(\hat{d}_n\).
Its convolved Gaussian likelihood is
\begin{equation}
    p_n(v)
    =
    \int_{v-\frac12}^{v+\frac12}
    \mathcal{N}\!\left(t;0,\sigma_n^2\right)\,dt.
    \label{eq:skip_residual_likelihood}
\end{equation}
At inference time, \(p_n(\hat{d}_n)\) gives the residual-symbol probability mass.
Small predicted scales imply that residuals are likely to round to zero.
For \(d\sim\mathcal{N}(0,\sigma^2)\) with unit-step rounding,
\begin{equation}
\small
    P(\mathrm{round}(d)=0)
    =
    P(-0.5 \leq d < 0.5) 
    =
    2\Phi\left(\frac{0.5}{\sigma}\right)-1,
\label{eq:zero_round_probability}
\end{equation}
where \(\Phi(\cdot)\) denotes the cumulative distribution function of the standard normal distribution.

We therefore skip residuals whose decoder-available scale is below a threshold:
\begin{equation}
    s_n=\mathbb{I}\!\left(\sigma_n\leq\tau_{\sigma}\right),
    \label{eq:skip_indicator}
\end{equation}
where \(\tau_{\sigma}\) is the skip threshold.
Because \(s_n\) depends only on the decoder-available scale estimate \(\sigma_n\), encoder and decoder make the same decision before residual decoding; rules that inspect \(d_n\) or \(\hat{d}_n\) are oracle diagnostics only (Section~\ref{sec:entropy_skip_analysis}).

If \(s_n=1\), no residual symbol is transmitted and decoder sets it to zero; otherwise, residual is entropy coded normally:
\begin{equation}
    \tilde{d}_n=(1-s_n)\hat{d}_n,\qquad
    \tilde{y}_n=\mu_n+\tilde{d}_n .
    \label{eq:skip_final_quant}
\end{equation}
The decision does not depend on $d_n$ or $\hat{d}_n$; it only uses $\sigma_n$, which is available at both encoder and decoder before residual decoding.
Only non-skipped residual symbols enter the bitstream, and the same skip decisions place decoded symbols back while leaving skipped positions as zero residuals.
Since the skip mask is derived only from decoder-available scale estimates, it does not require additional signaling and preserves encoder--decoder synchronization.
This usage is similar to entropy skip in \cite{alphavc}, where highly predictable symbols are omitted in a decoder-synchronized manner.

During training, skipped elements are masked out from the primary-latent likelihood loss, matching the zero emitted rate in the bitstream and reducing the noise-relaxation mismatch for low-scale residuals.
For non-skipped elements, we use
\begin{equation}
    \bar{d}_n=d_n+u_n,\qquad
    u_n\sim\mathcal{U}\!\left(-\tfrac12,\tfrac12\right),
    \label{eq:skip_noise}
\end{equation}
and compute the skip-aware primary-latent rate as
\begin{equation}
    \mathcal{L}_{\mathrm{rate}}^{y,\mathrm{skip}}
    =
    -\sum_n (1-s_n)\log_2 p_n(\bar{d}_n).
    \label{eq:skip_rate_loss}
\end{equation}
The hyper-latent rate is computed normally and is unaffected by this mask.

\subsection{Two-Stage Rate-Distortion Optimization}
\label{subsec:loss}

The codec is trained with learned rate terms plus spectral and adversarial distortions, without VQ, codebook, or commitment losses:
\begin{equation}\label{loss}
\small
\begin{aligned}
    \mathcal{L}_{\mathrm{total}}
    &=
    \mathcal{L}_{\mathrm{rate}}^{y,\mathrm{skip}}
    + \mathcal{L}_{\mathrm{rate}}^{z}
    + \lambda_{\mathrm{rd}}\mathcal{D},\\
    \mathcal{D}
    &=
    \lambda_{\mathrm{mel}}\mathcal{L}_{\mathrm{mel}}
    + \lambda_{\mathrm{adv}}\mathcal{L}_{\mathrm{Adv}}
    + \lambda_{\mathrm{fm}}\mathcal{L}_{\mathrm{FM}}
    + \lambda_{\mathrm{wav}}\mathcal{L}_{\mathrm{wav}} .
\end{aligned}
\end{equation}
Here $\mathcal{L}_{\mathrm{rate}}^{y,\mathrm{skip}}$ is defined in Eq.~\eqref{eq:skip_rate_loss}, $\mathcal{L}_{\mathrm{rate}}^{z}$ is the factorized hyper-latent rate, and $\mathcal{L}_{\mathrm{wav}}=\|x-\hat{x}\|_1$ is enabled only for objective-metric-oriented fine-tuning.

\subsubsection{Training Schedule}

Training proceeds in two stages.
Stage 1 trains a high-rate perceptual model with $\lambda_{\mathrm{rd}}=10$, disables entropy skip and waveform L1, and uses mel, adversarial, and feature-matching losses with MPD and MS-STFT discriminators.
Stage 2 fine-tunes rate-specific models from high to low bitrates, adjusts $\lambda_{\mathrm{rd}}$ for each target rate, enables entropy skip, and adds waveform L1 to improve objective reconstruction quality.

\subsubsection{Reconstruction Losses}

The main reconstruction term is a multi-scale mel-spectrogram loss, which captures spectral structure at different temporal resolutions:
\begin{align}\label{rec}
    \mathcal{L}_{\mathrm{mel}}
    &=
    \sum_{a\in\mathcal{A}}
    \big(
    \| \mathcal{S}_a(x) - \mathcal{S}_a(\hat{x}) \|_1 \nonumber\\
    &+ \beta
    \| \log \mathcal{S}_a(x) - \log \mathcal{S}_a(\hat{x}) \|_2
    \big),
\end{align}
where $\mathcal{S}_a$ denotes the mel-spectrogram transform at scale $a$, $\mathcal{A}$ is the set of spectral scales, and $\beta$ is a fixed log-magnitude weight.
The optional waveform loss $\mathcal{L}_{\mathrm{wav}}$ is used only when optimizing models for objective metrics.

\subsubsection{Adversarial and Feature-Matching Losses}

We use adversarial discriminators inspired by DAC~\cite{dac}.
MPD captures periodic waveform structure, while MS-STFT operates on multi-resolution complex spectra.
Let $\mathcal{K}$ denote the active discriminator set, which is stage-dependent as described above.
For a discriminator $D_k\in\mathcal{K}$, the generator adversarial loss and feature matching loss are
\begin{align}
    \mathcal{L}_{\mathrm{Adv}}
    &= \sum_{D_k\in\mathcal{K}} \mathbb{E}\!\left[-D_k(\hat{x})\right], \\
    \mathcal{L}_{\mathrm{FM}}
    &= \sum_{D_k\in\mathcal{K}} \sum_l
    \frac{1}{N_l}
    \left\| D_k^{(l)}(x)-D_k^{(l)}(\hat{x}) \right\|_1 ,
\end{align}
where $D_k^{(l)}$ is the feature map of the $l$-th discriminator layer and $N_l$ is the number of elements in that feature map.
Feature matching stabilizes adversarial training and encourages reconstructed speech to match the intermediate perceptual statistics of real speech.

\begin{table}[t!]
\centering
\caption{Key hyperparameters and constants.}
\label{tab:train_hparams}
\renewcommand{\arraystretch}{1.05}
\setlength{\tabcolsep}{4pt}
\footnotesize
\begin{tabular}{@{}>{\raggedright\arraybackslash}p{0.14\linewidth}
                >{\raggedright\arraybackslash}p{0.39\linewidth}
                >{\raggedright\arraybackslash}p{0.39\linewidth}@{}}
\toprule
\textbf{Block} & \textbf{Category} & \textbf{Setting} \\
\midrule

Transform 
& multi-scale codec stages $N$ & $4$ \\
module & linear-atten layers (per stage) & $\{2,4,6,8\}$ \\
& Embedding dim (per stage) & $\{1024,512,256,128\}$ \\
& Primary latent channels $C_y$ & $320$ \\
\midrule

Entropy & Hyper-latent channels $C_z$ & $192$ \\
module & channel slices $S$ & $5$ \\
& Channels per slice & $64$ \\
\midrule

Training & $\lambda_{\mathrm{mel}}, \lambda_{\mathrm{adv}}, \lambda_{\mathrm{fm}}$
& $1,\; 1/9,\; 100/9$ \\
objective & Stage-1 $\lambda_{\mathrm{rd}}$ & $10$ \\
& Stage-2 $\lambda_{\mathrm{rd}}$ & target-dependent \\
& Multi-scale spectral set $\mathcal{A}$ & $\{5,6,\dots,11\}$ \\
& STFT/Mel window for scale $i$ & $2^i$ \\
& STFT/Mel hop for scale $i$ & $2^i/4$ \\
\bottomrule
\end{tabular}
\vspace{-5mm}
\end{table}

%% file: sections/experiment.tex
\section{Experiment}
\label{sec:experiment}

\begin{figure*}[htbp]
    \centering
    \includegraphics[width=1 \textwidth]{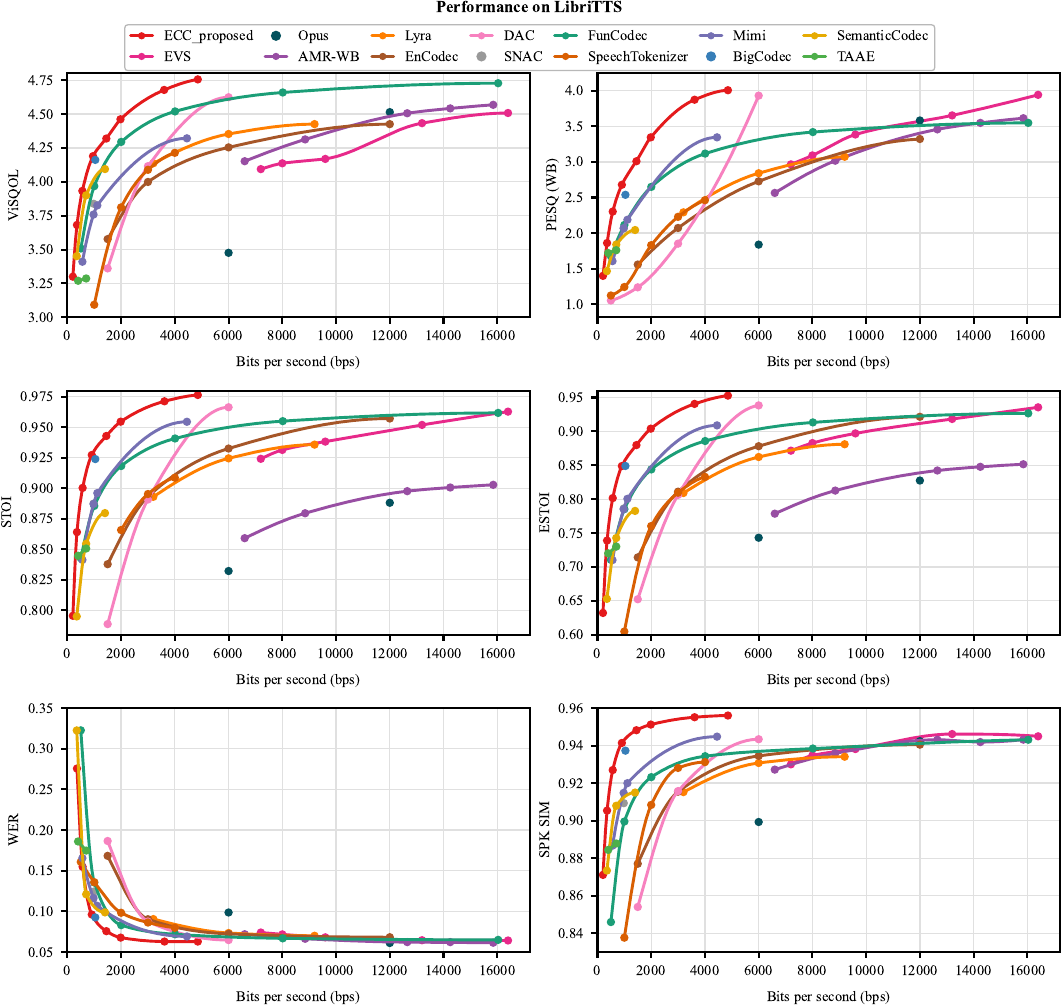}
    \vspace{-7mm}
    \caption{RD performance on LibriTTS across the objective metric set.
    \textbf{ECC} shows a strong low-bitrate RD trade-off.}
    \label{fig:libritts_rd_curves}
\vspace{-5mm}
\end{figure*}

\providecommand{\bdBetter}{\cellcolor{green!12}}
\providecommand{\bdWorse}{\cellcolor{red!12}}

\begin{table*}[htbp]
\centering
\caption{BD comparison relative to FunCodec on LibriTTS test-all.}
\label{tab:libritts_bd_benchmark}
\scriptsize
\setlength{\tabcolsep}{3.0pt}
\resizebox{\textwidth}{!}{%
\begin{tabular}{@{}llrrrrrrrrrrrr@{}}
\toprule
\textbf{Group} &
\textbf{Method} &
\multicolumn{6}{c}{\textbf{BD-rate $\downarrow$}} &
\multicolumn{6}{c}{\textbf{BD-metric}} \\
\cmidrule(lr){3-8}\cmidrule(l){9-14}
& & \textbf{ViSQOL} & \textbf{PESQ} & \textbf{STOI} & \textbf{ESTOI} & \textbf{WER} & \textbf{SPK-SIM}
& \textbf{ViSQOL $\uparrow$} & \textbf{PESQ $\uparrow$} & \textbf{STOI $\uparrow$} & \textbf{ESTOI $\uparrow$} & \textbf{WER $\downarrow$} & \textbf{SPK-SIM $\uparrow$} \\
\midrule
\multirow{3}{*}{\textbf{Classic}} & EVS~\cite{evs} & \bdWorse 435.39\% & \bdBetter \underline{-18.70\%} & \bdWorse 118.91\% & \bdWorse 55.21\% & \bdWorse 49.30\% & \bdWorse 35.12\% & \bdWorse -0.404 & \bdWorse -0.018 & \bdWorse -0.0146 & \bdWorse -0.0149 & \bdWorse 0.0014 & \bdBetter 0.0004 \\
& Opus~\cite{opus} & \bdWorse 644.32\% & \bdWorse 253.10\% & \bdWorse 1195.10\% & \bdWorse 738.67\% & \bdWorse 152.35\% & \bdWorse 279.95\% & \bdWorse -0.669 & \bdWorse -0.718 & \bdWorse -0.0954 & \bdWorse -0.1290 & \bdWorse 0.0132 & \bdWorse -0.0180 \\
& AMR-WB~\cite{amr} & \bdWorse 254.53\% & \bdWorse 50.94\% & \bdWorse 900.67\% & \bdWorse 561.82\% & \bdBetter -19.13\% & \bdWorse 74.83\% & \bdWorse -0.244 & \bdWorse -0.154 & \bdWorse -0.0721 & \bdWorse -0.0959 & \bdBetter -0.0010 & \bdWorse -0.0021 \\
\midrule
\multirow{9}{*}{\textbf{Neural}} & FunCodec~\cite{funcodec} & 0\% & 0\% & 0\% & 0\% & 0\% & 0\% & 0 & 0 & 0 & 0 & 0 & 0 \\
& SoundStream~\cite{soundstream} & \bdWorse 143.09\% & \bdWorse 135.22\% & \bdWorse 158.19\% & \bdWorse 131.29\% & \bdWorse 81.14\% & \bdWorse 101.15\% & \bdWorse -0.271 & \bdWorse -0.513 & \bdWorse -0.0286 & \bdWorse -0.0460 & \bdWorse 0.0073 & \bdWorse -0.0084 \\
& EnCodec~\cite{encodec} & \bdWorse 199.17\% & \bdWorse 186.84\% & \bdWorse 141.01\% & \bdWorse 110.71\% & \bdWorse 74.84\% & \bdWorse 88.64\% & \bdWorse -0.407 & \bdWorse -0.676 & \bdWorse -0.0302 & \bdWorse -0.0453 & \bdWorse 0.0169 & \bdWorse -0.0115 \\
& DAC~\cite{dac} & \bdWorse 289.34\% & \bdWorse 77.19\% & \bdWorse 586.87\% & \bdWorse 332.64\% & \bdWorse 180.31\% & \bdWorse 241.24\% & \bdWorse -0.724 & \bdWorse -0.850 & \bdWorse -0.1030 & \bdWorse -0.1445 & \bdWorse 0.1360 & \bdWorse -0.0607 \\
& SpeechTokenizer~\cite{speechtokenizer} & \bdWorse 293.86\% & \bdWorse 214.85\% & \bdWorse 615.59\% & \bdWorse 403.86\% & \bdWorse 11.35\% & \bdWorse 132.93\% & \bdWorse -0.729 & \bdWorse -0.776 & \bdWorse -0.0961 & \bdWorse -0.1429 & \bdBetter -0.0139 & \bdWorse -0.0423 \\
& Mimi~\cite{moshi} & \bdWorse 51.77\% & \bdBetter -2.94\% & \bdBetter \underline{-9.75\%} & \bdBetter \underline{-7.39\%} & \bdBetter -15.10\% & \bdBetter -41.76\% & \bdWorse -0.210 & \bdBetter 0.019 & \bdBetter \underline{0.0054} & \bdBetter 0.0070 & \bdBetter -0.0196 & \bdBetter 0.0149 \\
& SemantiCodec~\cite{semanticodec} & \bdBetter \underline{-22.46\%} & \bdWorse 6.10\% & \bdWorse 24.77\% & \bdWorse 8.48\% & \bdBetter \textbf{-33.76\%} & \bdBetter \underline{-45.47\%} & \bdBetter \underline{0.127} & \bdWorse -0.077 & \bdWorse -0.0147 & \bdWorse -0.0123 & \bdBetter \underline{-0.0781} & \bdBetter 0.0312 \\
& TAAE~\cite{taae} & \bdWorse 49.15\% & \bdBetter -5.37\% & \bdBetter -1.37\% & \bdBetter -6.22\% & \bdBetter -30.94\% & \bdBetter -32.41\% & \bdWorse -0.273 & \bdBetter \underline{0.035} & \bdBetter 0.0011 & \bdBetter \underline{0.0078} & \bdBetter \textbf{-0.1198} & \bdBetter \underline{0.0343} \\
& \textbf{ECC} & \bdBetter \textbf{-44.19\%} & \bdBetter \textbf{-69.38\%} & \bdBetter \textbf{-58.35\%} & \bdBetter \textbf{-55.67\%} & \bdBetter \underline{-32.06\%} & \bdBetter \textbf{-80.36\%} & \bdBetter \textbf{0.268} & \bdBetter \textbf{0.597} & \bdBetter \textbf{0.0386} & \bdBetter \textbf{0.0625} & \bdBetter -0.0594 & \bdBetter \textbf{0.0515} \\
\bottomrule
\end{tabular}
}

\vspace{1mm}
\begin{minipage}{\textwidth}
\scriptsize
\emph{Note:} FunCodec is the BD anchor. Green and red backgrounds indicate better and worse values than FunCodec according to each column direction. Bold and underline denote the best and second-best values in each column.
BD values are reported only for methods with more than two valid operating points within the common RD range.

\end{minipage}
\vspace{-3mm}
\end{table*}

\begin{figure*}[htbp]
    \centering
    \includegraphics[width=1 \textwidth]{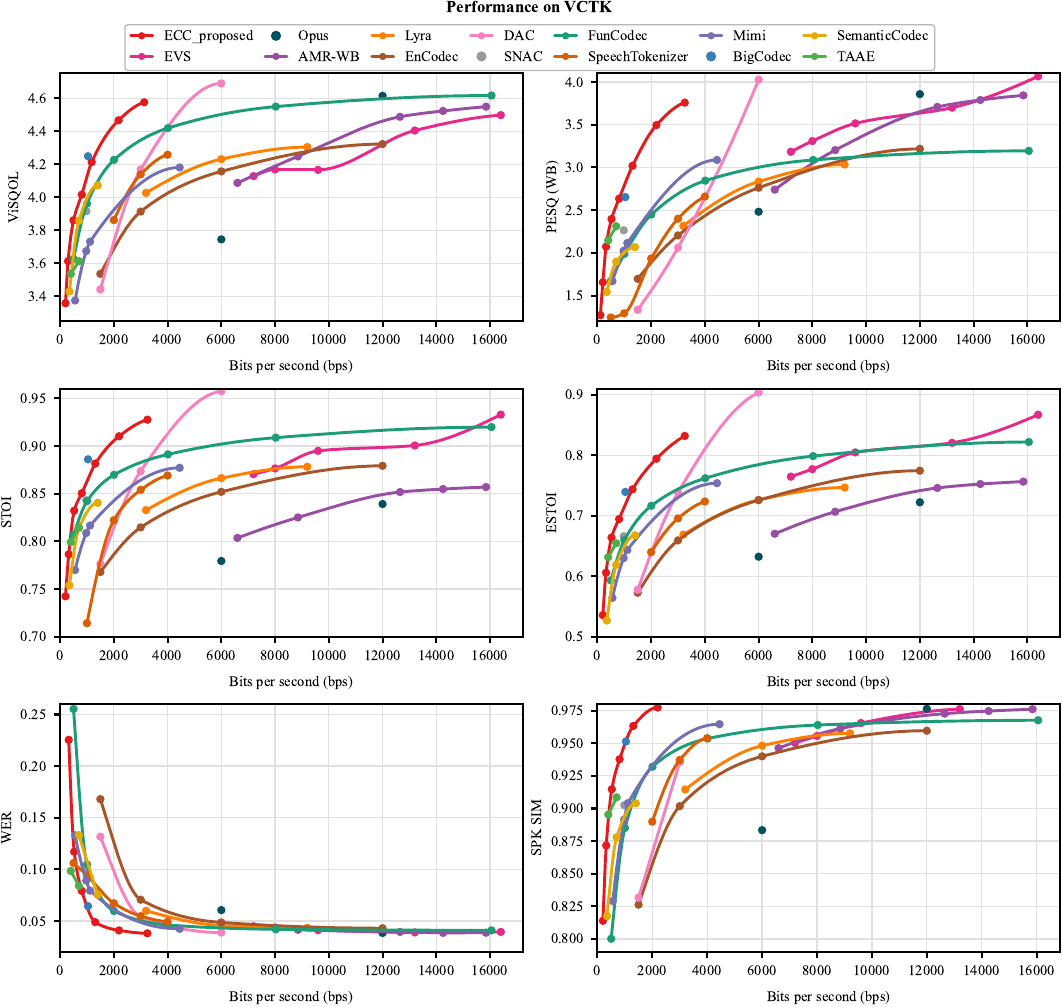}
    \vspace{-7mm}
    \caption{Rate--distortion performance on VCTK across the objective metric set.
    \textbf{ECC} preserves a favorable RD trade-off on the out-of-domain evaluation set.}
    \label{fig:vctk_rd_curves}
\vspace{-5mm}
\end{figure*}

\begin{table*}[htbp]
\centering
\caption{BD comparison relative to FunCodec on VCTK.}
\label{tab:vctk_bd_benchmark}
\scriptsize
\setlength{\tabcolsep}{3.0pt}
\resizebox{\textwidth}{!}{%
\begin{tabular}{@{}llrrrrrrrrrrrr@{}}
\toprule
\textbf{Group} &
\textbf{Method} &
\multicolumn{6}{c}{\textbf{BD-rate $\downarrow$}} &
\multicolumn{6}{c}{\textbf{BD-metric}} \\
\cmidrule(lr){3-8}\cmidrule(l){9-14}
& & \textbf{ViSQOL} & \textbf{PESQ} & \textbf{STOI} & \textbf{ESTOI} & \textbf{WER} & \textbf{SPK-SIM}
& \textbf{ViSQOL $\uparrow$} & \textbf{PESQ $\uparrow$} & \textbf{STOI $\uparrow$} & \textbf{ESTOI $\uparrow$} & \textbf{WER $\downarrow$} & \textbf{SPK-SIM $\uparrow$} \\
\midrule
\multirow{3}{*}{\textbf{Classic}} & EVS~\cite{evs} & \bdWorse 290.77\% & \bdBetter \textbf{-95.65\%} & \bdWorse 72.39\% & \bdBetter -18.35\% & \bdBetter -12.74\% & \bdBetter -33.58\% & \bdWorse -0.300 & \bdBetter 0.443 & \bdWorse -0.0177 & \bdBetter 0.0003 & \bdBetter -0.0006 & \bdBetter 0.0026 \\
& Opus~\cite{opus} & \bdWorse 319.75\% & \bdBetter -63.11\% & \bdWorse 1517.27\% & \bdWorse 583.76\% & \bdWorse 96.63\% & \bdWorse 208.75\% & \bdWorse -0.374 & \bdBetter 0.075 & \bdWorse -0.1003 & \bdWorse -0.1232 & \bdWorse 0.0076 & \bdWorse -0.0343 \\
& AMR-WB~\cite{amr} & \bdWorse 239.79\% & \bdBetter -74.01\% & \bdWorse 1090.88\% & \bdWorse 382.65\% & \bdBetter \underline{-28.77\%} & \bdWorse 1.38\% & \bdWorse -0.233 & \bdBetter 0.256 & \bdWorse -0.0780 & \bdWorse -0.0865 & \bdBetter -0.0007 & \bdWorse -0.0006 \\
\midrule
\multirow{9}{*}{\textbf{Neural}} & FunCodec~\cite{funcodec} & 0\% & 0\% & 0\% & 0\% & 0\% & 0\% & 0 & 0 & 0 & 0 & 0 & 0 \\
& SoundStream~\cite{soundstream} & \bdWorse 183.60\% & \bdWorse 62.77\% & \bdWorse 243.87\% & \bdWorse 165.70\% & \bdWorse 47.72\% & \bdWorse 95.90\% & \bdWorse -0.285 & \bdWorse -0.216 & \bdWorse -0.0390 & \bdWorse -0.0629 & \bdWorse 0.0042 & \bdWorse -0.0167 \\
& EnCodec~\cite{encodec} & \bdWorse 252.04\% & \bdWorse 85.44\% & \bdWorse 414.87\% & \bdWorse 192.97\% & \bdWorse 101.87\% & \bdWorse 144.39\% & \bdWorse -0.399 & \bdWorse -0.336 & \bdWorse -0.0604 & \bdWorse -0.0741 & \bdWorse 0.0222 & \bdWorse -0.0368 \\
& DAC~\cite{dac} & \bdWorse 609.27\% & \bdBetter -36.76\% & \bdWorse 391.35\% & \bdWorse 109.69\% & \bdWorse 279.59\% & \bdWorse 98.23\% & \bdWorse -0.701 & \bdWorse -0.550 & \bdWorse -0.0737 & \bdWorse -0.0900 & \bdWorse 0.1298 & \bdWorse -0.0671 \\
& SpeechTokenizer~\cite{speechtokenizer} & \bdWorse 509.20\% & \bdWorse 142.64\% & \bdWorse 566.39\% & \bdWorse 346.48\% & \bdWorse 3.55\% & \bdWorse 118.97\% & \bdWorse -0.718 & \bdWorse -0.514 & \bdWorse -0.0888 & \bdWorse -0.1345 & \bdBetter -0.0201 & \bdWorse -0.0754 \\
& Mimi~\cite{moshi} & \bdWorse 97.28\% & \bdBetter -12.80\% & \bdWorse 91.51\% & \bdWorse 34.85\% & \bdBetter -16.90\% & \bdBetter -11.19\% & \bdWorse -0.278 & \bdBetter 0.071 & \bdWorse -0.0273 & \bdWorse -0.0239 & \bdBetter -0.0172 & \bdBetter 0.0096 \\
& SemantiCodec~\cite{semanticodec} & \bdBetter \underline{-8.06\%} & \bdBetter -13.34\% & \bdWorse 38.87\% & \bdWorse 15.54\% & \bdWorse 0.77\% & \bdBetter -28.13\% & \bdBetter \underline{0.034} & \bdBetter 0.042 & \bdWorse -0.0156 & \bdWorse -0.0135 & \bdBetter -0.0082 & \bdBetter 0.0310 \\
& TAAE~\cite{taae} & \bdWorse 17.58\% & \bdBetter -63.65\% & \bdWorse \underline{4.99\%} & \bdBetter \underline{-36.57\%} & \bdBetter \textbf{-53.72\%} & \bdBetter \underline{-56.72\%} & \bdWorse -0.082 & \bdBetter \underline{0.540} & \bdWorse \underline{-0.0026} & \bdBetter \underline{0.0442} & \bdBetter \textbf{-0.1466} & \bdBetter \textbf{0.0933} \\
& \textbf{ECC} & \bdBetter \textbf{-35.65\%} & \bdBetter \underline{-83.25\%} & \bdBetter \textbf{-8.44\%} & \bdBetter \textbf{-43.60\%} & \bdBetter -11.14\% & \bdBetter \textbf{-69.99\%} & \bdBetter \textbf{0.186} & \bdBetter \textbf{0.693} & \bdBetter \textbf{0.0096} & \bdBetter \textbf{0.0507} & \bdBetter \underline{-0.0418} & \bdBetter \underline{0.0892} \\
\bottomrule
\end{tabular}%
}

\vspace{1mm}
\begin{minipage}{\textwidth}
\scriptsize
\emph{Note:} FunCodec is the BD anchor. Green and red backgrounds indicate better and worse values than FunCodec according to each column direction. Bold and underline denote the best and second-best values in each column.
BD values are reported only for methods with more than two valid operating points within the common RD range.
\end{minipage}
\vspace{-3mm}
\end{table*}

\subsection{Experimental Setup}

\subsubsection{Dataset}

We train ECC on LibriTTS~\cite{libritts}.
For the main objective comparison, we evaluate all codecs on LibriTTS test-all, which is the union of test-clean and test-other, and on the VCTK~\cite{vctk} test set.
LibriTTS test-all serves as the in-domain evaluation set, while VCTK provides an English out-of-domain evaluation with different speakers and recording conditions.
We further use AISHELL-3~\cite{aishell3} as a Mandarin Chinese test set to examine cross-lingual generalization beyond the English training corpus.

\subsubsection{Training Details}

ECC is trained with the two-stage objective in Section~\ref{subsec:loss}: a high-rate perceptual model is first trained and then fine-tuned from high to low RD operating points.
The main comparison uses $\tau_{\sigma}=0.12$ unless otherwise specified.
Table~\ref{tab:train_hparams} summarizes the key settings; no VQ, codebook, or commitment loss is used.

\subsubsection{Baselines}

We compare \textbf{ECC} with conventional codecs, OPUS~\cite{opus}, EVS~\cite{evs}, and AMR~\cite{amr}, and with neural codecs listed below.
Table~\ref{tab:codec_summary} provides the broader taxonomy across transform domains, quantizers, and model families.
Baseline results use open-source implementations and released weights when available.
The neural baselines include SoundStream~\cite{soundstream}\footnote{\url{https://github.com/google/lyra}}, 
EnCodec~\cite{encodec}\footnote{\url{https://github.com/facebookresearch/encodec}},
DAC~\cite{dac}\footnote{\url{https://github.com/descriptinc/descript-audio-codec}},
SNAC~\cite{snac}\footnote{\url{https://github.com/hubertsiuzdak/snac}},
FunCodec~\cite{funcodec}\footnote{\url{https://github.com/modelscope/FunCodec}},
SpeechTokenizer~\cite{speechtokenizer}\footnote{\url{https://github.com/ZhangXInFD/SpeechTokenizer}},
Mimi~\cite{moshi}\footnote{\url{https://github.com/kyutai-labs/moshi}},
BigCodec~\cite{bigcodec}\footnote{\url{https://github.com/Aria-K-Alethia/BigCodec}},
SemantiCodec~\cite{semanticodec}\footnote{\url{https://github.com/haoheliu/SemantiCodec-inference}},
and TAAE~\cite{taae}\footnote{\url{https://github.com/Stability-AI/stable-codec}}.
For neural baselines, we use publicly released implementations and pretrained checkpoints whenever available, and follow the official configurations to generate operating points.
All decoded waveforms use the same preprocessing and metric pipeline.
Since codecs differ in operating range and training corpus, this comparison evaluates publicly available codec systems rather than a controlled architecture-only setting.

\begin{figure*}[htbp]
    \centering
    \includegraphics[width=1 \textwidth]{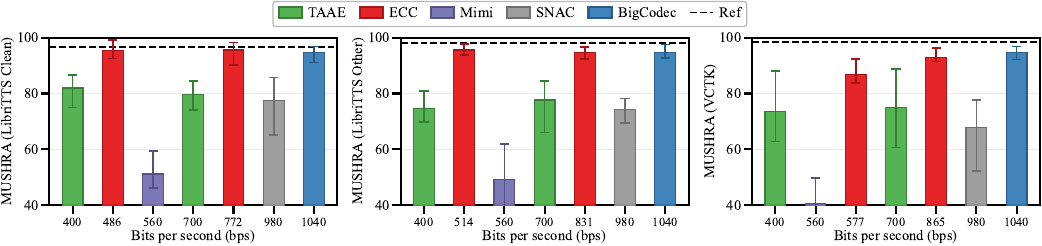}
    \vspace{-7mm}
    \caption{MUSHRA subjective results in the low-bitrate regime.
    Bars and error bars denote mean listener scores and standard deviations; \textbf{ECC} achieves strong perceived quality at lower bitrates.}
    \label{fig:mushra}
\vspace{-2mm}
\end{figure*}

\begin{figure*}[htp]
    \centering
    \begin{minipage}[t]{0.66\textwidth}
        \centering
        \includegraphics[width=\linewidth]{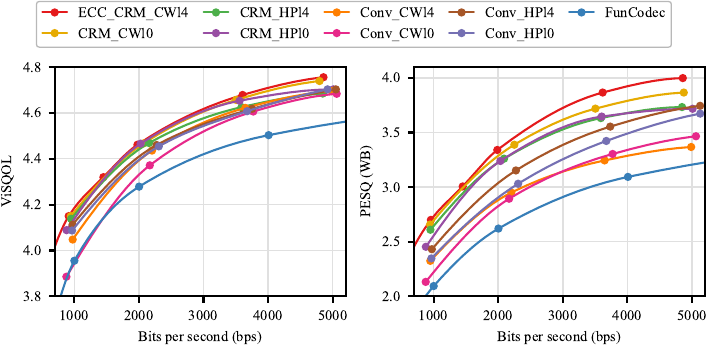}
    \end{minipage}\hfill
    \begin{minipage}[t]{0.32\textwidth}
        \centering
        \includegraphics[trim=0mm 0mm 0mm 0mm, clip, width=\linewidth]{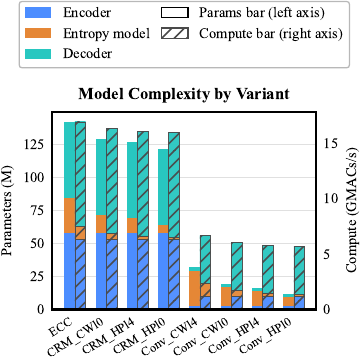}
    \end{minipage}
    \vspace{-3mm}
    \caption{Ablation and complexity results.
    Left: ablation study on LibriTTS test-all using ViSQOL and PESQ, comparing backbone design, entropy structure, and entropy attention depth.
    Right: complexity comparison among variants.}
    \label{fig:ablation}
    \label{fig:complexity}
\vspace{-5mm}
\end{figure*}

\subsubsection{Evaluation Metrics}

We report ViSQOL~\cite{visqol}, wideband PESQ~\cite{pesq}, STOI~\cite{stoi}, ESTOI~\cite{estoi}, WER, speaker similarity, and actual bitrate.
For ECC, the reported bitrate is computed from the actual entropy-coded bitstream, including both hyper-latent and primary-latent streams.
For neural RVQ baselines, rates follow the official operating points or the generated coded representations provided by the released systems.

For metric computation, all reference waveforms are resampled to 16~kHz, and reconstructed waveforms are evaluated against the corresponding 16~kHz references.
No loudness normalization, amplitude normalization, or additional time-domain alignment is applied.
WER is computed with a HuBERT-based ASR backend~\cite{hubert} by comparing the recognized text with the ground-truth transcript of each dataset.
Speaker similarity is computed with a WavLM-based speaker verification backend~\cite{wavlm}.
All methods are evaluated with the same preprocessing and metric backends.
For codecs with sparse released operating points, intermediate RD samples are interpolated for curve-level comparison, while BD-rate and BD-metric values are reported only for methods with more than two valid points in the common quality range.

\subsection{Rate-Distortion Performance}
\label{subsec:rd_performance}

\subsubsection{Objective RD Curves}

Figs.~\ref{fig:libritts_rd_curves} and~\ref{fig:vctk_rd_curves} report objective RD curves on LibriTTS test-all and VCTK, respectively.
Across perceptual quality, intelligibility, recognition, and speaker-preservation metrics, ECC consistently occupies a favorable low-bitrate region compared with both conventional codecs and recent neural baselines.
On LibriTTS, ECC achieves comparable or better quality at lower actual bitrates, showing the effectiveness of entropy-constrained scalar-latent coding on the in-domain test set.
On VCTK, ECC preserves a similar trend under speaker and recording-condition shifts, indicating that the learned representation and entropy model generalize beyond the training corpus.

\begin{table}[tbp]
\centering
\caption{BD ablation results of the variants.}
\label{tab:ablation}
\setlength{\tabcolsep}{3.0pt}
\begin{tabular}{@{}lcccc@{}}
\toprule
\textbf{Variant} &
\makecell{\textbf{BD-ViSQOL $\uparrow$}} &
\makecell{\textbf{BD-rate $\downarrow$}\\\textbf{(ViSQOL)}} &
\makecell{\textbf{BD-PESQ $\uparrow$}} &
\makecell{\textbf{BD-rate $\downarrow$}\\\textbf{(PESQ)}} \\
\midrule
ECC\_CRM\_CWl4 & \textbf{0.195} & \textbf{-45.86\%} & \textbf{0.737} & \textbf{-65.60\%} \\
CRM\_CWl0 & \underline{0.189} & \underline{-44.98\%} & \underline{0.673} & \underline{-64.18\%} \\
CRM\_HPl4 & 0.167 & -40.30\% & 0.597 & -60.64\% \\
CRM\_HPl0 & 0.184 & -43.26\% & 0.580 & -58.75\% \\
Conv\_CWl4 & 0.124 & -31.98\% & 0.248 & -31.09\% \\
Conv\_CWl0 & 0.065 & -18.04\% & 0.210 & -27.48\% \\
Conv\_HPl4 & 0.141 & -36.46\% & 0.444 & -49.34\% \\
Conv\_HPl0 & 0.130 & -34.21\% & 0.328 & -40.55\% \\
\bottomrule
\end{tabular}

\vspace{1mm}
\begin{minipage}{\linewidth}
\scriptsize
\emph{Note:} CRM/Conv denote the CRM and purely convolutional encoder--decoder backbones. CW/HP compare channel-wise context modeling with LRP against hyperprior-only entropy modeling. l0/l4 denote zero or four attention layers in the entropy model.
\end{minipage}
\vspace{-6mm}
\end{table}

Tables~\ref{tab:libritts_bd_benchmark} and~\ref{tab:vctk_bd_benchmark} further summarize curve-level performance using FunCodec as the anchor.
ECC achieves the best or near-best BD-rate and BD-metric results for most reported metrics on both datasets, with particularly consistent gains in perceptual quality and speaker similarity.
The recognition metric is more competitive across methods, but ECC remains among the leading approaches while maintaining strong perceptual and speaker-preservation performance.
Overall, the objective results show that ECC improves the low-bitrate RD trade-off under both in-domain and out-of-domain evaluations.

\subsubsection{Subjective Test}

\begin{figure*}[t]
    \centering
    \includegraphics[trim=0mm 0mm 0mm 0mm, clip, width=\textwidth]{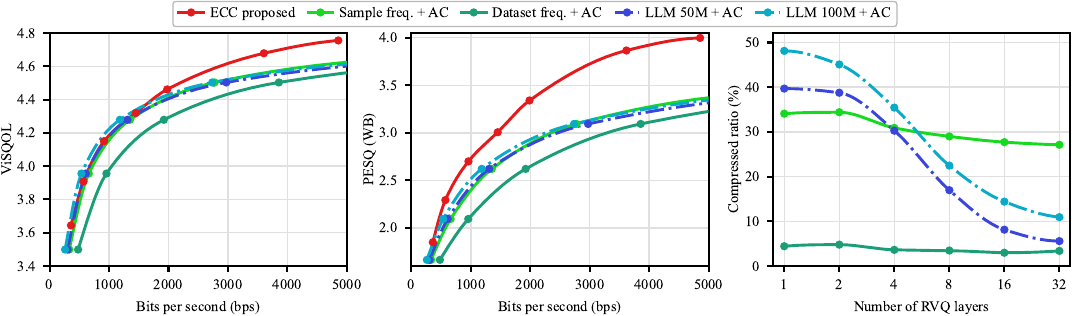}
    \vspace{-7mm}
    \caption{Post-hoc entropy-coding diagnostics.
    Left: comparison between \textbf{ECC} and post-hoc entropy coding baselines; FunCodec variants re-encode fixed RVQ indices, so quality is unchanged and only bitrate shifts.
    Right: per-RVQ-stage compression ratio on FunCodec indices; higher values indicate stronger lossless compression.}
    \label{fig:posthoc_entropy_compare}
    \label{fig:ac_compression_ratio}
\vspace{-3mm}
\end{figure*}

\begin{figure*}[t]
    \centering
    \includegraphics[trim=0mm 0mm 0mm 0mm, clip, width=\textwidth]{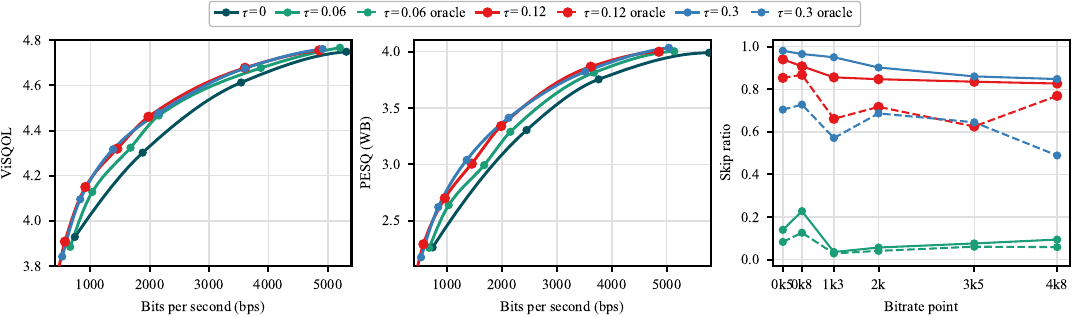}
    \vspace{-7mm}
    \caption{Entropy skip threshold analysis.
    Left: rate--distortion comparison of entropy skip thresholds; larger thresholds skip more residual symbols, and $\tau_{\sigma}=0.12$ is used in the main comparison.
    Right: skip ratio statistics for the normal scale-threshold rule and the oracle diagnostic rule.}
    \label{fig:ablation_skip}
    \label{fig:skip_ratio}
\vspace{-5mm}
\end{figure*}

We further assess low-bitrate perceptual quality with a MUSHRA listening test.
The test includes reference and low-quality anchor samples, and compares ECC with representative low-bitrate neural codec baselines.
As shown in Fig.~\ref{fig:mushra}, ECC obtains high subjective scores on both LibriTTS test-clean/test-other and VCTK utterances.
It remains close to the strongest competing neural codec while operating at a lower bitrate, and outperforms several baselines in the same low-bitrate range.
These subjective results are consistent with the objective RD curves and confirm that the entropy-constrained representation improves perceived speech quality at very low bitrates.

\subsection{Ablation Studies}
\label{subsec:ablation_studies}

We analyze the main design choices of ECC, including the transform backbone, channel-wise entropy modeling, entropy-side temporal modeling, post-hoc coding of fixed RVQ indices, and entropy skip.

\subsubsection{Architecture Ablation} 

Table~\ref{tab:ablation} and Fig.~\ref{fig:ablation} compare three architectural factors: CRM versus purely convolutional backbones, channel-wise context with LRP versus hyperprior-only entropy modeling, and four versus zero RWKV layers in the entropy model.
Across the tested RD range, CRM variants consistently outperform their convolutional counterparts under comparable entropy settings, showing the benefit of hybrid local and long-range spectro-temporal modeling. 
Under the CRM backbone, channel-wise context modeling with LRP further improves the RD trade-off over the hyperprior-only setting, indicating that decoded channel context captures dependencies not fully explained by the hyperprior.
Adding entropy-side temporal modeling generally improves the PESQ-oriented trade-off, while the l0 variants show that much of the gain already comes from the transform backbone and channel-wise entropy structure.
Overall, the best configuration combines the CRM transform, channel-wise context with LRP, and lightweight temporal modeling in the entropy model.

\begin{figure*}[t]
    \centering
    \begin{minipage}[t]{0.66\textwidth}
        \centering
        \includegraphics[width=\linewidth]{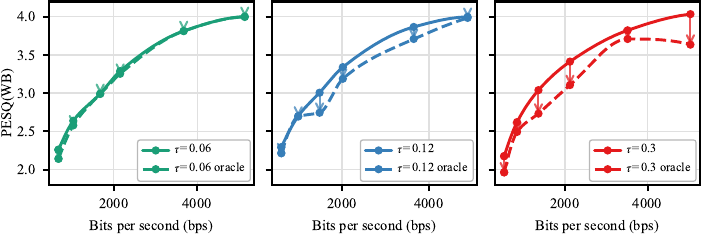}
    \end{minipage}\hfill
    \begin{minipage}[t]{0.33\textwidth}
        \centering
        \includegraphics[trim=0mm 0mm 0mm 0mm, clip, width=\linewidth]{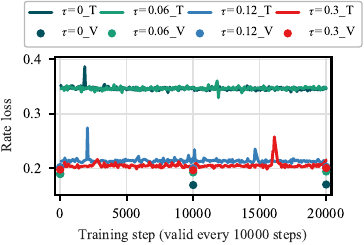} 
    \end{minipage}
    \vspace{-3mm}
    \caption{Entropy skip diagnostics.
    Left: diagnostic PESQ comparison between normal skip and oracle skip for $\tau_{\sigma}=0.06$, $0.12$, and $0.3$.
    Right: training and validation latent rate loss under different skip thresholds; larger thresholds reduce the train--validation gap.}
    \label{fig:oracle_skip_rd}
    \label{fig:rate_loss}
\vspace{-3mm}
\end{figure*}

\begin{figure*}[htbp]
    \centering
    \begin{minipage}[t]{0.66\textwidth}
        \centering
        \includegraphics[trim=0mm 0mm 0mm 0mm, clip, width=\linewidth]{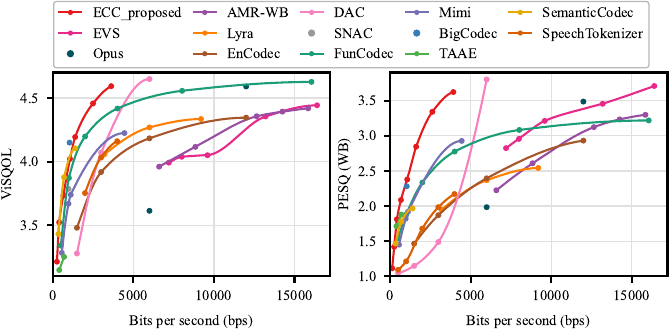}
    \end{minipage}\hfill
    \begin{minipage}[t]{0.33\textwidth}
        \centering
        \includegraphics[trim=0mm 0mm 0mm 0mm, clip, width=\linewidth]{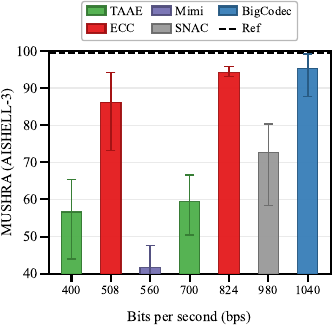}
    \end{minipage}
    \vspace{-3mm}
    \caption{Generalization performance on AISHELL-3.
    \textbf{ECC} maintains a favorable low-bitrate RD trade-off on Mandarin Chinese speech, suggesting cross-lingual generalization beyond the LibriTTS training corpus.}
    \label{fig:generalization_aishell3}
\vspace{-5mm}
\end{figure*}

\subsubsection{Post-Hoc Coding Versus Learned Latents}
\label{sec:posthoc_entropy_compare}

To compare post-hoc index coding with joint entropy-constrained learning, we re-encode fixed FunCodec RVQ indices using dataset-level marginal, sample-level marginal, and autoregressive Transformer coders.
These baselines only change the lossless coding of an already generated index stream, so waveform quality and objective distortion metrics remain unchanged.
As shown in Fig.~\ref{fig:posthoc_entropy_compare}, post-hoc coding can reduce bitrate, but the gain depends strongly on the coding model and the RVQ stage.
Early RVQ stages contain stronger temporal regularities and are more compressible, whereas later residual stages become harder to predict.
Thus, post-hoc coding can reduce the cost of a fixed representation, but it cannot reshape the latents, codebook usage, or reconstruction behavior.
ECC instead learns scalar-quantized latents jointly with their probability model, yielding a stronger low-bitrate RD trade-off.

\begin{table*}[htbp]
\centering
\caption{Complexity comparison with neural codec baselines.}
\label{tab:baseline_complexity}
\setlength{\tabcolsep}{1.4pt}
\begin{tabular}{@{}lcccccccccc@{}}
\toprule
\textbf{Metric} &
\textbf{EnCodec\cite{encodec}} &
\textbf{DAC\cite{dac}} &
\textbf{SNAC\cite{snac}} &
\textbf{FunCodec\cite{funcodec}} &
\textbf{SpeechTokenizer\cite{speechtokenizer}} &
\textbf{Mimi\cite{moshi}} &
\textbf{BigCodec\cite{bigcodec}} &
\textbf{SemanticCodec\cite{semanticodec}} &
\textbf{TAAE\cite{taae}} &
\textbf{ECC} \\
\midrule
\textbf{GMACs/s} & \underline{5.56} & 55.65 & 7.278 & \textbf{2.143} & 17.045 & 11.214 & 61.092 & 1077.779 & 37.568 & 16.930 \\
\textbf{Params(M)} & \underline{14.85} & 74.06 & 19.84 & \textbf{4.5} & 103.676 & 79.292 & 159.323 & 507 & 953.09 & 150.68 \\
\bottomrule
\end{tabular}
\vspace{-3mm}
\end{table*}

\subsubsection{Entropy Skip Thresholds and Coding Consistency}
\label{sec:entropy_skip_analysis}

We evaluate entropy skip thresholds $\tau_{\sigma}\in\{0,0.06,0.12,0.3\}$, where $\tau_{\sigma}=0$ denotes the no-skip baseline and $\tau_{\sigma}=0.12$ is used in the main comparison.
As shown in Fig.~\ref{fig:ablation_skip}, skip-enabled models improve over the no-skip baseline, with larger thresholds producing higher skip ratios and stronger RD gains.
The threshold $\tau_{\sigma}=0.12$ achieves most of the gain without using the most aggressive skip ratio, and is therefore adopted as a conservative main setting.
The normal skip rule is deployable because it depends only on decoder-available scale estimates, while oracle skip is used only as a diagnostic because it depends on the actual rounded residual.

Fig.~\ref{fig:oracle_skip_rd} compares normal skip with oracle skip and reports the training--validation rate behavior.
The oracle skip ratio indicates how many skipped positions are truly zero after rounding, while the gap between normal and oracle ratios shows that the scale-threshold rule can also suppress some nonzero residuals.
Although oracle skip restores these nonzero rounded residuals, it does not necessarily improve RD performance because later context prediction and LRP are trained under the scale-threshold skip trajectory.
Recovering these residuals can therefore introduce a mismatch in the decoded latent trajectory.
The rate-loss curves further show that entropy skip reduces the gap between noise-relaxed training rates and rounded-symbol validation rates, especially at larger thresholds.
These results indicate that entropy skip improves coding efficiency by omitting highly predictable residual symbols and by making the training rate term more consistent with actual coding.

\subsection{Complexity}
\label{subsec:complexity}

Table~\ref{tab:baseline_complexity} reports the computational complexity and parameter counts of ECC and representative neural codec baselines.
The reported GMACs/s and parameters measure the neural-network components of each codec, including the analysis transform, synthesis transform, and entropy-model networks when applicable.
They provide an architecture-level comparison rather than a hardware-specific runtime measurement, since practical latency also depends on implementation details, arithmetic coding, and sequential entropy-decoding steps.

Compared with compact codecs such as FunCodec~\cite{funcodec}, EnCodec~\cite{encodec}, SNAC~\cite{snac}, and Mimi~\cite{moshi}, ECC requires more parameters and computation due to its stronger transform backbone and explicit entropy model.
Nevertheless, its GMACs/s remains comparable to SpeechTokenizer~\cite{speechtokenizer} and lower than heavier systems such as DAC~\cite{dac}, BigCodec~\cite{bigcodec}, SemanticCodec~\cite{semanticodec}, and TAAE~\cite{taae}, reflecting a complexity--RD trade-off.

The complexity comparison also shows where future optimization is needed.
Channel-wise entropy modeling and slice-wise decoding improve coding efficiency, but they may introduce extra sequential operations during deployment.
Lightweight entropy models, faster context prediction, and streaming-oriented implementations are therefore important directions for practical low-delay speech coding.

\subsection{Generalization}
\label{subsec:generalization}

We evaluate the proposed ECC, trained on the English LibriTTS speech dataset, using the AISHELL-3 dataset to examine its cross‑lingual generalization to Mandarin Chinese speech.
Since ECC is trained as a waveform reconstruction codec rather than a language model, this experiment mainly tests whether the learned acoustic representation and entropy model transfer beyond the English training corpus.
As shown in Fig.~\ref{fig:generalization_aishell3}, ECC maintains a favorable low-bitrate RD trade-off on both ViSQOL and PESQ.
Its curves rise quickly in the low-to-mid bitrate range, indicating that the entropy-constrained representation remains effective under language and recording-condition shifts.
Some baselines approach competitive quality only at substantially higher bitrates, whereas ECC achieves strong perceptual quality with fewer transmitted bits. 

These results suggest that ECC does not simply overfit to the LibriTTS test distribution.
Instead, the learned scalar latents and probability model retain useful acoustic compression behavior on a cross-lingual test set.
Broader multilingual and general-audio evaluations remain important future work.

%% file: sections/conclusion.tex
\section{Conclusion}

In this work, we benchmarked neural speech compression from a rate--distortion perspective and studied entropy-constrained coding as a way to improve low-bitrate efficiency.
We formulated a unified learning-based speech coding pipeline, reviewed recent neural speech codecs, and identified the mismatch between preset-rate discrete representations and learned probability modeling.
To address this issue, we proposed ECC, an Entropy-Constrained Codec that combines scalar quantization, hyperprior-based side information, channel-wise context modeling, lightweight temporal modeling, latent residual prediction, and entropy skip within an end-to-end rate--distortion optimization framework.
Extensive experiments show that ECC achieves a favorable low-bitrate RD trade-off under objective and subjective evaluations, with 44.2\%/35.7\% ViSQOL and 69.4\%/83.3\% PESQ BD-rate reductions over FunCodec on LibriTTS/VCTK datasets.
Ablation and diagnostic results further validate the effectiveness of entropy modeling, context prediction, post-hoc coding analysis, and skip-aware rate optimization.
Future work includes practical rate-control mechanisms for constant- or adaptive-bitrate deployment, lightweight and low-latency entropy modeling, and broader evaluations on multilingual, noisy-speech, and general-audio scenarios.